\documentclass[showpacs,floatfix,superscriptaddress]{revtex4}
\usepackage{graphicx}
\usepackage{epsfig}
\usepackage{bm}
\usepackage[utf8]{inputenc}
\usepackage{amssymb}
\usepackage{float}
\usepackage{amsmath}
\usepackage{dcolumn}
\usepackage{latexsym}
\usepackage{subfig}
\usepackage{amsthm}
\usepackage{framed}
\usepackage{cancel}
\usepackage{mathtools}
\newcommand\myeq{\mathrel{\stackrel{\makebox[0pt]{\mbox{\normalfont\tiny $\tilde{R}\rightarrow\infty$}}}{=}}}
\newcommand\myeqq{\mathrel{\stackrel{\makebox[0pt]{\mbox{\normalfont\tiny $\tilde{R}\rightarrow\infty$}}}{\rightarrow}}}
\usepackage[colorlinks]{hyperref}
\usepackage[dvipsnames]{xcolor}
\hypersetup{
     breaklinks=true,
    pdfstartview={FitH},    
    colorlinks=true,       
    linkcolor=blue,          
    citecolor=red,        
    filecolor=magenta,      
    urlcolor=blue,           
    anchorcolor=green,      
    linktocpage=true
}

\def\doi{http://doi.org}

\begin{document}

\title{Asymptotically locally flat and AdS higher-dimensional black holes of Einstein-Horndeski-Maxwell gravity in the light of EHT observations: shadow behavior and deflection angle}

\author{Kourosh Nozari}
\email[]{knozari@umz.ac.ir}
\affiliation{Department of Theoretical Physics, Faculty of Sciences, University of Mazandaran,\\
47416-95447, Babolsar, Iran}

\author{Sara Saghafi}
\email[]{s.saghafi@umz.ac.ir}
\affiliation{Department of Theoretical Physics, Faculty of Sciences, University of Mazandaran,\\
47416-95447, Babolsar, Iran}

\begin{abstract}
Unification of gravity with other interactions, achieving the ultimate framework of quantum gravity, and fundamental problems in particle physics and cosmology motivate to consider extra spatial dimensions. The impact of these extra dimensions on the modified theories of gravity has attracted a lot of attention. One way to examine how extra dimensions affect the modified gravitational theories is to analytically investigate astrophysical phenomena, such as black hole shadows. In this study, we aim to investigate the behavior of the shadow shapes of higher-dimensional charged black hole solutions including asymptotically locally flat (ALF) and asymptotically locally AdS (ALAdS) in Einstein-Horndeski-Maxwell (EHM) gravitational theory. We utilize the Hamilton-Jacobi method to find photon orbits around these black holes as well as the Carter approach to formulate the geodesic equations. We examine how extra dimensions, negative cosmological constant, electric charge, and coupling constants of the EHM gravity affect the shadow size of the black hole. Then, we constrain these parameters by comparing the shadow radius of these black holes with the shadow size of M87* supermassive black hole captured by the Event Horizon Telescope (EHT) collaborations. We discover that generally the presence of extra dimensions within the EHM gravity results in reducing the shadow size of higher-dimensional ALF and ALAdS charged black holes, whereas the impact of electric charge on the shadow of these black holes is suppressible. Interestingly, we observe that decreasing the negative cosmological constant, i.e., increasing its absolute value, leads to increase the shadow size of the ALAdS charged higher-dimensional black hole in the EHM gravity. Surprisingly, based on the constraints from EHT observations, we discover that only the shadow size of the four dimensional ALF charged black hole lies in the confidence levels of EHT data, whereas owing to the presence of the negative cosmological constant, the shadow radius of the four, five, and seven dimensional ALAdS charged black holes lie within the EHT data confidence levels.
\vspace{12 pt}
\end{abstract}


\maketitle

\enlargethispage{\baselineskip}
\tableofcontents

\section{Introduction}

The breakthrough successes in capturing the first images of shadows of the supermassive black holes M87* \cite{EventHorizonTelescope:2019dse,EventHorizonTelescope:2019uob,EventHorizonTelescope:2019jan,EventHorizonTelescope:2019ths,EventHorizonTelescope:2019pgp,
EventHorizonTelescope:2019ggy,EventHorizonTelescope:2021bee,EventHorizonTelescope:2021srq} and Sgr A* \cite{EventHorizonTelescope:2022wkp,EventHorizonTelescope:2022apq,EventHorizonTelescope:2022wok,EventHorizonTelescope:2022exc,EventHorizonTelescope:2022urf,
EventHorizonTelescope:2022xqj} by the Event Horizon Telescope (EHT) collaboration shed light on the physics of black holes and open a wide gate to a deeper understanding of these mysterious celestial objects. The event horizon of black holes, i.e., the boundary of no return for any crossing matter or radiation, is not directly observable, since it emits no light. Instead, what we can observe is the black hole ``shadow'', which is the dark region on a light background that appears around the event horizon due to the gravitational lensing phenomenon \cite{Perlick:2004tq,Virbhadra:2022iiy,Virbhadra:2022ybp}. Since releasing the shadow images of M87* and Sgr A*, many efforts have been devoted to improving measurements to reach higher-resolution images \cite{Goddi:2016qax}. As a result, theoretical efforts in investigating black hole physics, particularly shadow behavior in various gravitational theories, become significant in producing the desired resolution. In this regard, analytic and numerical studies and examinations of the apparent geometrical shape of various black hole spacetimes supply new theoretical shadow templates for future observations \cite{Vagnozzi:2022moj,Bambi:2019tjh,Vagnozzi:2020quf,Roy:2021uye,Chen:2022nbb}. The shape and size of the shadow is determined by the black hole parameters, i.e., mass, electric charge, and angular momentum \cite{deVries:1999tiy} in addition to spacetime properties \cite{Johannsen:2010ru,Cunha:2017eoe} and the position of the observer. For non-rotating black holes, shape of the shadow is a perfect circle. The angular momentum parameter can, however, cause rotating black holes to have non-trivial shadow shapes. \cite{Chandrasekhar:1984siy}.

In 1914, Nordström first proposed the idea of extra dimensions \cite{Nordstrom:1914ejq}. According to his idea, one can unify the electromagnetic and gravitational fields by treating four-dimensional spacetime as a surface in a five-dimensional spacetime. Today, unifying gravitational and gauge interactions of elementary particles, quantizing gravitational interaction, the Higgs mass hierarchy problem, and the cosmological constant problem are the main motivations for the enormous quantity of studies on the extra dimensions. In this regard, the Kaluza-Klein (KK) theory \cite{Kaluza:1921tu,Klein:1926tv} utilizing Einstein's General Theory of Relativity (GR) introduces a compact space constructed by compact extra dimensions with a certain compactification scale to unify gravitational interaction and electromagnetic or even non-Abelian gauge fields characterizing weak and strong interactions. Furthermore, the string theory (M-theory) as the well-known candidate theory for quantum gravity possesses eleven compact extra spatial dimensions or more \cite{Witten:1995ex,Schwarz:1995jq}. In addition to these compact extra dimensions with the extension up to the order of the Planck length, there are also ideas for large extra dimensions to the order of millimeter. This new gate to the topic of extra dimensions has opened by the Arkani-Hamed-Dimopoulos-Dvali (ADD) braneworld model \cite{Arkani-Hamed:1998jmv,Arkani-Hamed:1998sfv} to address the Higgs mass hierarchy problem via employing large extra dimensions. It is worth noting that the dramatic feature of these large extra dimensions is that their impacts can be detectable in future accelerator, astrophysical and tabletop experiments. Surprisingly, the ADD model can be incorporated in string theory \cite{Antoniadis:1998ig}. Besides the compact and large extra dimensions, Randall-Sundrum (RS) braneworld model \cite{Randall:1999ee,Randall:1999vf} suggests the warped extra dimensions to address the Higgs mass hierarchy problem. In addition to these types of extra dimensions, there are also theories with infinite volume extra dimensions like Dvali-Gabadadze-Porrati (DGP) braneworld scenario \cite{Dvali:2000hr} in which even in very low energies, the spacetime is not four-dimensional and the extra dimensions are neither compact nor warped. Such theories are the candidates for addressing the cosmological constant problem \cite{Dvali:2000xg,Dvali:2002pe} since in these theories, gravity is modified at large distances thanks to presence of infinite volume extra dimensions. Some detailed reviews on higher-dimensional models can be seen in Refs. \cite{Gabadadze:2003ii,Shifman:2009df,Perez-Lorenzana:2005fzz,Rubakov:2001kp}. Within the framework of black hole physics, different methods and approaches have been employed to extend and investigate various black hole models in arbitrary dimensions \cite{Emparan:2008eg,Kanti:2004nr}, such as Tangherlini method \cite{Tangherlini:1963bw} for generalizing the Schwarzschild solution to $n$ dimensions.

Detecting extra dimensions is a priority for physicists in high-energy or particle experiments. The Large Hadron Collider (LHC) at CERN and future colliders become some promising tools for exploring such extra dimensions and effects of strong gravity regimes corresponding with higher-dimensional black holes \cite{Allanach:2002gn,Agashe:2006hk,Franceschini:2011wr,Deutschmann:2017bth,Strominger:1996sh,Harris:2004xt,Antoniadis:2000vd}. Moreover, the presence of Hydrogen atom in higher dimensions \cite{Burgbacher:1999sha,Caruso:2012daf,Shaqqor:2009cha}, spectroscopy experiments \cite{Zhou:2014xbw,Luo:2006ck,Luo:2006ad}, and the ideas to address the proton radius puzzle \cite{Wang:2013fma,Dahia:2015bza,Zhi-gang:2007swh} support the existence of extra dimensions. On the other hand, two recent achievements towards discovering black hole strong field regimes are the detection of Gravitational Waves (GW) via the LIGO/Virgo collaborations \cite{LIGOScientific:2016aoc}, and the above-mentioned images captured by the EHT collaborations. There are some traces of extra dimensions in the detection of GW, possessing certain information about the associate amplitude and the dynamics of fluctuation modes. Hence, many works have been focused on revealing such physics \cite{Cardoso:2016rao,Yu:2016tar,Visinelli:2017bny,Kwon:2019gsa,Rahman:2022fay} and for a detailed review, see Ref. \cite{Yu:2019jlb}. Now EHT has provided new possibilities to continue explorations for extra dimensions. Recently, in seminal works \cite{Vagnozzi:2019apd,Belhaj:2020mlv,Tang:2022hsu}, the authors found noteworthy constraints from EHT observations on warped and compact extra dimensions within the RS model and M-theory. Therefore, one can utilize the EHT data to explore all types of extra dimensions, generally, and see whether they can be detected, as we aim to do that in this study. In this regard, it seems that extra dimensions affect the shadows of black holes by reducing the shadow size in various black hole models and gravitational theories \cite{Amarilla:2011fx,Eiroa:2017uuq,Papnoi:2014aaa,Singh:2017vfr,Amir:2017slq,Belhaj:2020rdb}. However, the exact effect of extra dimensions on black hole shadows is still an area of active research and is not yet fully understood. Apparently, the impacts of large and infinite volume extra dimensions have more chance to be detected in the future.

Besides, the size and shape of black hole shadows (similar to other astrophysical phenomena \cite{Nozari:2020swx,Hajebrahimi:2020xvo,Nozari:2020tks,Nozari:2012nf,
Saghafi:2021wzx}) may differ in extended theories of gravity through additional degrees of freedom arose from these theories. Therefore, investigating the size and shape of the black hole shadows may aid in evaluating parameters of black hole metrics and testing alternative theories of gravity. Theoretical motivations \cite{Burgess:2003jk} or dark energy, dark matter, and cosmological modeling \cite{Clifton:2011jh,Nojiri:2006gh,Creminelli:2017sry,Sakstein:2017xjx,Ezquiaga:2017ekz,Heisenberg:2018vsk,Barack:2018yly} are only a few examples of the many hypotheses that make up the enormous field of extended theories of gravity beyond GR. Among them, some novel ghost-free special classes of theories have developed, such as $f(R)$ theories \cite{DeFelice:2010aj,Sotiriou:2008rp}, Lovelock theories \cite{Lovelock:1971yv,Lovelock:1972vz}, and the scalar-tensor theories initially formulated by Horndeski \cite{Horndeski:1974wa} (for detailed reviews, see Refs. \cite{Nojiri:2010wj,Nojiri:2017ncd}). The Horndeski theory is the most general scalar-tensor gravitational theory possessing second-order derivatives in the equations of motion. There are a lot of studies in the literature focused on the examining the Einstein-Horndeski scalar-tensor modified theory of gravity in various astrophysical issues and cosmological modeling \cite{Kobayashi:2019hrl,Volkova:2019jlj,Galeev:2021xit,
Frusciante:2018jzw,Starobinsky:2016kua,Babichev:2016rlq,Silva:2016smx,Maselli:2016gxk,Charmousis:2011bf,Salahshoor:2018plr}. To obtain the stable black hole solutions of Einstein-Horndeski gravity, it has received significant attention to consider the action containing a non-minimal kinetic coupling of one scalar field to the Einstein tensor field. Spherically symmetric solutions with non-minimal derivative coupling without cosmological constant has been investigated in Ref. \cite{Rinaldi:2012vy}, and considering a negative cosmological constant studied in Ref. \cite{Minamitsuji:2013ura}. The asymptotically locally flat and asymptotically locally anti de Sitter (AdS) black hole solutions in Einstein-Horndeski gravity were also first found in Ref. \cite{Anabalon:2013oea}. The asymptotically locally flat and asymptotically locally AdS black hole solutions in the Einstein-Horndeski-Maxwell (EHM) gravitational theory with four and extra dimensions were obtained in Ref. \cite{Cisterna:2014nua}. The thermodynamics of the later solution is also studied in Refs. \cite{Feng:2015wvb,Hajian:2020dcq}.

A vast number of works have been focused on the issue of black hole shadow to find what and how degrees of freedom, arose from extended theories of gravity other than black holes parameters, affect the shadow behavior \cite{Perlick:2021aok}. Some examples are as follows: the shadow behavior of the Kerr-Newman family of solutions of the Einstein-Maxwell equations is investigated in Refs. \cite{Bardeen:1973tla,Takahashi:2004xh,Ghosh:2022mka}; the shadow of a black hole with NUT-charges \cite{Chakraborty:2013kza,Grenzebach:2014fha}; the black hole shadows in Einstein-Maxwell-dilaton gravity \cite{Amarilla:2013sj,Wei:2013kza}, in Chern-Simons modified gravity \cite{Amarilla:2010zq}; the apparent shape of the Sen black hole \cite{Hioki:2008zw,Dastan:2016bfy,Younsi:2016azx}; shadows of colliding and multi-black holes \cite{Nitta:2011nin,Yumoto:2012kz}; shadow behavior of rotating black holes in $f(R)$ gravity \cite{Dastan:2016vhb}, conformal Weyl gravity \cite{Mureika:2016efo}, and Einstein-dilaton-Gauss-Bonnet black holes \cite{Cunha:2016wzk}; shadow behavior of the non-commutative geometry inspired, quantum-corrected, and magnetically charged black holes \cite{Wei:2015dua,Sharif:2016znpx,Saghafi:2022pmey,Allahyari:2019jqz}; shadow behavior of Einstein-Born-Infeld black holes \cite{Atamurotov:2015xfa}; shadow behavior of Ayon-Beato-Garcia black hole and also, rotating Hayward and rotating Bardeen regular black holes \cite{Abdujabbarov:2016hnw} and hairy black holes \cite{Cunha:2015yba,Cunha:2016bjh,Khodadi:2020jij}; chaotic shadow of a non-Kerr rotating compact objects with quadrupole mass moment and a magnetic dipole \cite{Wang:2018eui,Wang:2017qhh}, and black holes with exotic matter \cite{Tinchev:2015apf,Abdujabbarov:2015pqp,Singh:2017xle,Huang:2016qnl,Chowdhuri:2020ipb,
Sheikhahmadi:2023jpb}; and also, shadow behavior of wormholes and naked singularities \cite{Nedkova:2013msa,Ohgami:2015nra,Ortiz:2015rma}.

In this study, we aim to investigate the shadow behavior and deflection angle of the asymptotically locally flat (ALF) and asymptotically locally AdS (ALAdS) charged black hole solutions in EHM gravity with extra dimensions and also, estimate the energy emission rate associated with these black holes. We want to examine how extra dimensions together with electric charge and negative cosmological constant within the EHM gravity affect the shadow and deflection angle of the black holes to gain a new template of black hole shadow for future theoretical and observational applications. Additionally, we want to constrain extra dimensions, the electric charge, negative cosmological constant, and the coupling constants of EHM gravity by comparing the shadow size of the higher-dimensional ALF and ALAdS charged black holes in EHM gravity with the shadow size of M87* supermassive black hole captured by EHT. This paper is organized as follows. In Section \ref{EHM:bls} we first briefly introduce the EHM gravitational theory with arbitrary dimensions and then describe the line elements of the higher-dimensional ALF and ALAdS charged black holes in the theory. In Section \ref{genfor:hdbl:so}, we provide the general formalism to study the shadow behavior of the higher-dimensional black holes by utilizing the Hamilton-Jacobi approach and Carter method to formulate the null geodesic equations. We specify the shadow shape of the black holes on the observer's sky in celestial coordinates, and estimate the energy emission rate and deflection angle formulas in higher dimensions. Also, we introduce the black hole shadow observables. In Section \ref{shadef:alfads}, utilizing the framework introduced in the previous section, we study the shadow behavior, deflection angle, and energy emission rates of the ALF and ALAdS charged black holes in EHM gravity with extra dimensions. We analyze the significant impacts of the electric charge, cosmological constant, extra dimensions, and the coupling constants of EHM gravity on the shadow and deflection angle of the black holes within the setup and then, we constrain these parameters by EHT data. Finally, Section \ref{consum} is devoted to discussing and concluding our main results.

\section{EHM gravity with arbitrary dimensions and its black hole solutions}\label{EHM:bls}

The action of the higher-dimensional Einstein-Horndeski gravity, which is minimally coupled to a Maxwell field to construct EHM gravity with arbitrary dimensions, has the following form \cite{Cisterna:2014nua,Feng:2015wvb}
\begin{equation}\label{action1}
I=\frac{1}{16\pi}\int d^{n}x\sqrt{-\tilde{g}}\mathcal{L}\,,
\end{equation}
in which $n$ counts the number of spacetime dimensions, and the Lagrangian is to the form of
\begin{equation}\label{lagrangian1}
\mathcal{L}=R-2\Lambda-\frac{1}{4}F_{ab}F^{ab}-\frac{1}{2}\left(\alpha g^{ab}-\gamma G^{ab}\right)\partial_{a}\chi\,\partial_{b}\chi\,,
\end{equation}
where $\alpha$ and $\gamma$ are the coupling constants, $F_{ab}=\partial_{a}A_{b}-\partial_{b}A_{a}$ is the electromagnetic field strength with gauge potential $A$, and $G_{ab}\equiv R_{ab}-\frac{1}{2}R\,g_{ab}$ is the Einstein tensor in which $R_{ab}$ is the Ricci tensor, $R$ is Ricci scalar, and $g_{ab}$ is the metric tensor with determinant $\tilde{g}$. The Lagrangian possesses the derivatives of the axionic scalar field $\chi$. This makes the Lagrangian invariant under the transformation $\chi\rightarrow\chi+C$. Here, this symmetry, however, does not utilize to yield the non-minimally coupled Einstein-vector gravity \cite{Geng:2015kvs}. The strength of the non-minimal kinetic coupling to Einstein tensor field is governed by $\gamma$.

By varying the action \eqref{action1} with respect to the metric tensor, axionic scalar field, and the gauge potential, one can find the corresponding equations of motion in the EHM gravity \cite{Cisterna:2014nua,Feng:2015wvb}. In order to find the static charged black hole solutions of the setup, one can take into account the following general spherically symmetric ansatz with arbitrary dimensions as the line element (metric tensor) of the background spacetime
\begin{equation}\label{ansatz}
ds^{2}=-h(r)dt^{2}+\frac{dr^{2}}{f(r)}+r^{2}d\Omega_{n-2}^{2}\,,
\end{equation}
where $d\Omega_{n-2}^{2}=d\theta_{1}^{2}+\sin^{2}[\theta_{1}]d\theta_{2}^{2}+\ldots+\prod_{i=1}^{n-3}\sin^{2}[\theta_{i}]d\theta_{n-2}^{2}$ is the metric of the unit $S^{n-2}$ hypersphere, which has the volume
\begin{equation}\label{volhysph}
\omega_{n-2}=\frac{2\pi^{\frac{n-1}{2}}}{\Gamma[\frac{n-1}{2}]}\,,
\end{equation}
where $\Gamma$ is the gamma function. By this ansatz, one can solve the equations of motion in EHM gravity to obtain two classes of higher-dimensional black hole solutions, which are ALF and ALAdS black holes as constructed and reviewed in Refs. \cite{Cisterna:2014nua,Feng:2015wvb}.

\subsection{ALF black hole with extra dimensions in EHM gravity}

Setting $\alpha=\Lambda=0$ and also $\gamma<0$ (for a real scalar field outside the event horizon), the equations of motion of the EHM gravity result in the higher-dimensional ALF charged black hole solution for which we have (for more details, see Refs. \cite{Cisterna:2014nua,Feng:2015wvb})
\begin{equation}\label{fralf}
f(r)=\frac{16(n-2)^{2}(n-3)^{2}r^{4n}}{\left(q^{2}r^{6}-4(n-2)(n-3)r^{2n}\right)^{2}}h(r)\,,
\end{equation}
\begin{equation}\label{hralf}
h(r)=1-\frac{\mu}{r^{n-3}}+\frac{q^{2}}{2(n-2)(n-3)r^{2(n-3)}}-\frac{q^{4}}{48(n-2)^{2}(n-3)^{2}r^{4(n-3)}}\,,
\end{equation}
where $\mu$ and $q$ are two non-trivial parameters, which parameterise the mass and the electric charge, respectively in such a way that
\begin{equation}\label{masscharge1}
M=\frac{1}{16\pi}(n-2)\mu\,\omega_{n-2}\,,\qquad
Q=\frac{1}{8\pi}q\,\omega_{n-2}\sqrt{2(n-2)(n-3)}\,.
\end{equation}
It is worth noting that the parameter $q$ in the form as introduced in \cite{Feng:2015wvb} is not correct and we provided its correct form in Eq. \eqref{masscharge1}.

The higher-dimensional ALF charged black hole solution possesses two curvature singularities at $r=0$ and $r=r_{*}$, respectively, so that $r_{*}$ can be obtained through the following equation
\begin{equation}\label{singularity1}
4(n-2)(n-3)r_{*}^{2n-6}-q^{2}=0\,.
\end{equation}
On the other hand, the event horizon of the higher-dimensional ALF charged black hole is located at $r=r_{eh}$, which is the largest root of $h(r)=0$. Furthermore, the higher-dimensional ALF charged black hole satisfies the condition $r_{eh}>r_{*}$, which implies \cite{Cisterna:2014nua,Feng:2015wvb}
\begin{equation}\label{condition}
\frac{\mu}{q}>\frac{4}{3\sqrt{(n-2)(n-3)}}\,.
\end{equation}
The Hawking temperature associated with the higher-dimensional ALF charged black hole can be found as follows \cite{Feng:2015wvb}
\begin{equation}\label{Htemalf}
T_{ALF}=\frac{4(n-2)(n-3)r_{eh}^{2n-6}-q^{2}}{16\pi(n-2)r_{eh}^{2n-5}}\,.
\end{equation}
This temperature is always positive, i.e., $T_{ALF}>0$ due to the above-mentioned condition based on which the curvature singularity $r_{*}$ must be inside the event horizon $r_{eh}$. Therefore, the Hawking temperature of the higher-dimensional ALF charged black hole can approach zero, but can never reach this vanishing value. This feature is more in agreement with the behavior of physical systems respecting the third law of thermodynamics, and it cannot be seen in Reissner-Nordstr\"{o}m black hole.

\subsection{ALA\lowercase{d}S black hole with extra dimensions in EHM gravity}

Assuming $\alpha\neq 0$ and $\Lambda\neq 0$ so that $(\alpha+\gamma\Lambda)<0$ (to have a real scalar field outside the event horizon), the EHM gravity field equations result in the higher-dimensional ALAdS charged black hole solution for which we have (for more details, see Refs. \cite{Cisterna:2014nua,Feng:2015wvb})
\begin{equation}\label{fralads}
f(r)=\frac{(n-2)^{2}(4+\beta\gamma)^{2}\left((n-1)g^{2}r^{2}+n-3\right)^{2}}{\left((n-2)(n-1)(4+\beta\gamma)g^{2}r^{2}+4(n-2)(n-3)-q^{2}r^{2(3-n)}\right)^{2}}h(r)\,,
\end{equation}
\begin{equation}\label{hralads}
h(r)=\bar{h}(r)+h_{q}(r)\,,
\end{equation}
where
\begin{equation}\label{hrq}
\begin{split}
h_{q}(r) & =\frac{2q^{2}}{(n-2)(n-3)(4+\beta\gamma)r^{2n-6}}-\frac{2\beta\gamma(n-3)q^{2}}{g^{2}(n-1)^{2}(n-2)(4+\beta\gamma)^{2}r^{2n-4}}\\
& +\frac{2\beta\gamma(n-3)^{2}q^{2}}{g^{4}(n+1)(n-1)^{2}(n-2)(4+\beta\gamma)^{2}r^{2n-2}}{}_{2}\mathrm{F}_{1}\left[1,\frac{n+1}{2};\frac{n+3}{2};\frac{3-n}{(n-1)g^{2}r^{2}}
\right]\\
& -\frac{q^{4}}{g^{2}(n-1)(n-2)^{2}(3n-7)(4+\beta\gamma)^{2}r^{2(2n-5)}}{}_{2}\mathrm{F}_{1}\left[1,\frac{3n-7}{2};\frac{3n-5}{2};\frac{3-n}{(n-1)g^{2}r^{2}}\right]\,
\end{split}
\end{equation}
where ${}_{2}\mathrm{F}_{1}$ is the hypergeometric function, which is well-defined for $n\geq 4$. Furthermore, when the dimension number $n$ is even, the function $\bar{h}(r)$ is to the following form
\begin{equation}\label{hbarreven}
\bar{h}_{\mathrm{even}}(r)=-\frac{\mu}{r^{n-3}}+\frac{8g^{2}r^{2}(2+\beta\gamma)+16}{(4+\beta\gamma)^{2}}+\frac{\beta^{2}\gamma^{2}g^{2}r^{2}}{(4+\beta\gamma)^{2}}
{}_{2}\mathrm{F}_{1}\left[1,\frac{1-n}{2};\frac{3-n}{2};\frac{3-n}{(n-1)g^{2}r^{2}}\right]\,.
\end{equation}
The function $\bar{h}_{even}(r)$ is divergent for odd integers of dimension number. When $n$ is odd ($n\geq 5$), the function $\bar{h}(r)$ is as follows
\begin{equation}\label{hbarrodd}
\bar{h}_{\mathrm{odd}}(r)=-\frac{\mu}{r^{n-3}}+\frac{8g^{2}r^{2}(2+\beta\gamma)+16}{(4+\beta\gamma)^{2}}+\frac{(n-1)\beta^{2}\gamma^{2}g^{4}r^{4}}{(n-3)(4+\beta\gamma)^{2}}
{}_{2}\mathrm{F}_{1}\left[1,\frac{n+1}{2};\frac{n+3}{2};\frac{(n-1)g^{2}r^{2}}{3-n}\right]\,.
\end{equation}
$\alpha$ and $\gamma$ must possess the same sign to achieve the ALAdS spacetime \cite{Cisterna:2014nua,Feng:2015wvb}. In Eqs. \eqref{fralads}-\eqref{hbarrodd} two parameters $g$ and $\beta$ are substituted for $\alpha$ and the cosmological constant $\Lambda$ so that
\begin{equation}\label{gbeta}
\alpha=\frac{1}{2}(n-1)(n-2)g^{2}\gamma\,,\qquad
\Lambda=-\frac{1}{4}(n-1)(n-2)g^{2}(2+\beta\gamma)\,.
\end{equation}
Again, $\mu$ and $q$ parameterise the mass and the electric charge as follows
\begin{equation}\label{masscharge2}
M=\frac{1}{64\pi}(n-2)(4+\beta\gamma)\mu\,\omega_{n-2}\,,\qquad
Q=\frac{1}{8\pi}q\,\omega_{n-2}\sqrt{2(n-2)(n-3)}\,.
\end{equation}

The higher-dimensional ALAdS charged black hole solution has two curvature singularities at $r=0$ and $r=r_{*}$, respectively. The curvature singularity $r_{*}$ is the roots of
\begin{equation}\label{singularity2}
(n-2)(n-1)(4+\beta\gamma)g^{2}r^{2}+4(n-2)(n-3)-q^{2}r^{2(3-n)}=0\,
\end{equation}
and located within the event horizon of the black hole $r_{eh}$, which is a root of $h(r)=0$. Moreover, the Hawking temperature of the higher-dimensional ALAdS charged black hole can be found as follows \cite{Feng:2015wvb}
\begin{equation}\label{Htemalads}
T_{ALAdS}=\frac{(n-1)g^{2}r_{eh}}{4\pi}+\frac{4(n-2)(n-3)r_{eh}^{2n-6}-q^{2}}{4\pi(n-2)(4+\beta\gamma)r_{eh}^{2n-5}}\,.
\end{equation}

\section{General formalism for shadow and deflection angle of higher-dimensional black holes and shadow observables}\label{genfor:hdbl:so}

When a black hole is in front of a light source, part of the light is deflected by the gravitational field of black hole and reaches the observer. However, some photons may fall into the black hole, creating a dark zone known as the shadow, and the apparent shape of the black hole is the boundary of the shadow. In this section, we present the general formulas required to obtain the shape of the shadow, energy emission rate, and deflection angle for the general ansatz \eqref{ansatz} with higher dimensions, which necessitates the study of the motion of a test particle in the spacetime.

\subsection{Null geodesics}

We start with the Lagrangian of the test particle, which is to the form of
\begin{equation}\label{lagrangian}
\tilde{\mathcal{L}}=\frac{1}{2}g_{ab}\dot{x}^{a}\dot{x}^{b}\,,
\end{equation}
where an over dot shows the derivative with respect to the affine parameter $\tau$. The components of canonically conjugate momentum corresponding with the general ansatz \eqref{ansatz} can be found as follows
\begin{equation}\label{pt}
P_{t}=h(r)\dot{t}=E\,,
\end{equation}
\begin{equation}\label{pr}
P_{r}=\frac{1}{f(r)}\dot{r}\,,
\end{equation}
\begin{equation}\label{pte}
P_{\theta_{i}}=r^{2}\sum_{i=1}^{n-3}\prod_{n=1}^{i-1}\sin^{2}[\theta_{n}]\dot{\theta}_{i}\,,
\end{equation}
\begin{equation}\label{pph}
P_{\theta_{n-2}}=r^{2}\prod_{i=1}^{n-3}\sin^{2}[\theta_{i}]\dot{\theta}_{n-2}=L\,,
\end{equation}
where $i=1,2,\cdots,n-3$ and also, $E$ and $L$ are the energy and angular momentum of the test particle, respectively.

We utilize the Hamilton-Jacobi method to analyze photon orbits around the black hole, in addition to the Carter approach to investigate the geodesic equations \cite{Carter:1968rr}. In this regard, we generalize these methods to higher dimensions. Consequently, in higher dimensions, the Hamilton-Jacobi method reads
\begin{equation}\label{HJ1}
\frac{\partial S}{\partial\tau}=-\frac{1}{2}g^{ab}\frac{\partial S}{\partial x^{a}}\frac{\partial S}{\partial x^{b}}\,,
\end{equation}
where $S$ is the Jacobi action of the test particle. Inserting the general ansatz \eqref{ansatz} with arbitrary dimensions into Eq. \eqref{HJ1}, one can yield
\begin{equation}\label{HJ2}
\begin{split}
-2 \frac{\partial S}{\partial\tau}=-\frac{1}{h(r)}\left(\frac{\partial S_t}{\partial t}\right)^{2}+f(r)\left(\frac{\partial S_r}{\partial r}\right)^{2}+\sum_{i=1}^{n-3}\frac{1}{\left(r^{2}\prod_{n=1}^{i-1}\sin^{2}[\theta_{n}]\right)}\left(\frac{\partial S_{\theta_{i}}}{\partial \theta_{i}}\right)^{2}+\frac{1}{\left(r^{2}\prod_{i=1}^{n-3}\sin^{2}[\theta_{i}]\right)}\left(\frac{\partial S_{\theta_{n-2}}}{\partial\theta_{n-2}}\right)^{2}\,.
\end{split}
\end{equation}

Taking into account a separable solution for Jacobi action allows one to express the action as
\begin{equation}\label{jaction}
S=\frac{1}{2} m^{2}\tau-Et+L\theta_{n-2}+S_{r}(r)+\sum^{n-3}_{i=1}S_{\theta_{i}}(\theta_{i})\,,
\end{equation}
where $m$ is the rest mass of the test particle. Since in studying shadow behavior of black holes, the test particle is photon, we set $m=0$. Therefore, applying the Jacobi action \eqref{jaction} on Eq. \eqref{HJ2} results in the following expression
\begin{equation}\label{HJ3}
\begin{split}
0 & =\left\{\frac{E^{2}}{h(r)}-f(r)\left(\frac{\partial S_r}{\partial r}\right)^{2}-\frac{1}{r^{2}}\left(\frac{L^{2}}{\prod_{i=1}^{n-3}\sin^{2}[\theta_{i}]}+\mathcal{K}-\prod_{i=1}^{n-3}L^{2}\cot^{2}[\theta_{i}]\right)\right\}\\
& -\left\{\frac{1}{r^{2}}\left(\sum_{i=1}^{n-3}\frac{1}{\prod_{n=1}^{i-1}\sin^{2}[\theta_{n}]}
\left(\frac{\partial S_{\theta_{i}}}{\partial \theta_{i}}\right)^{2}-\mathcal{K}+\prod_{i=1}^{n-3}L^{2}\cot^{2}[\theta_{i}]\right)\right\}\,,
\end{split}
\end{equation}
where $\mathcal{K}$ is the Carter constant. After some manipulations, one can obtain the following set of equations
\begin{equation}\label{HJ4}
r^{4}f^{2}(r)\left(\frac{\partial S_r}{\partial r}\right)^{2}=r^{4}\frac{f(r)}{h(r)}E^{2}-r^{2}\left(L^{2}+\mathcal{K}\right)f(r)\,,
\end{equation}
\begin{equation}\label{HJ5}
\sum_{i=1}^{n-3}\frac{1}{\prod_{n=1}^{i-1}\sin^{2}[\theta_{n}]}\left(\frac{\partial S_{\theta_{i}}}{\partial \theta_{i}}\right)^{2}=\mathcal{K}-\prod_{i=1}^{n-3}L^{2}\cot^{2}[\theta_{i}]\,.
\end{equation}
Finally, employing Eqs. \eqref{HJ4} and \eqref{HJ5} and the components of the canonically conjugate momentum \eqref{pt}-\eqref{pph}, the complete equations of motion for photon, i.e., the null geodesics within the higher-dimensional spacetime \eqref{ansatz} can be read as follows
\begin{equation}\label{td}
\dot{t}=\frac{E}{f(r)}\,,
\end{equation}
\begin{equation}\label{rd}
r^{2}\dot{r}=\pm\sqrt{\mathcal{R}}\,,
\end{equation}
\begin{equation}\label{ted}
r^{2}\sum_{i=1}^{n-3}\prod_{n=1}^{i-1}\sin^{2}[\theta_{n}]\dot{\theta}_{i}=\pm\sqrt{\Theta_{i}}\,,
\end{equation}
\begin{equation}\label{phd}
\dot{\theta}_{n-2}=\frac{L}{r^{2}\prod_{i=1}^{n-3}\sin^{2}[\theta_{i}]}\,,
\end{equation}
where ``$+$'' and ``$-$'' signs denote the outgoing and ingoing radial directions of the motion of photon, respectively. Furthermore, we have
\begin{equation}\label{RT}
\mathcal{R}=r^{4}\frac{f(r)}{h(r)}E^{2}-r^{2}\left(L^{2}+\mathcal{K}\right)f(r)\,,\quad \Theta_{i}=\mathcal{K}-\prod_{i=1}^{n-3}L^{2}\cot^{2}[\theta_{i}]\,.
\end{equation}
The motion of photon in the spacetime is governed by Eqs. \eqref{td}-\eqref{phd}.

It is critical to discuss the effective potential for determining the boundary of the shadow of black holes. The effective potential can be calculated by rewriting the radial null geodesic equation \eqref{rd} as follows
\begin{equation}\label{orbits}
\left(\frac{dr}{d\tau}\right)^{2}+V_{eff}=0\,,
\end{equation}
in which the effective potential is to the following form
\begin{equation}\label{veff}
V_{eff}=\frac{f(r)}{r^{2}}\left(\mathcal{K}+L^{2}\right)-\frac{f(r)}{h(r)}E^{2}\,.
\end{equation}

The unstable circular orbits of photons determine the boundary of apparent shape of the black hole. They correspond with the maximum value of the effective potential, which occurs at a distance, known as photon sphere radius $r_{0}$ satisfying the following equations
\begin{equation}\label{photonra}
V_{eff}\big|_{r_{0}}=\frac{dV_{eff}}{dr}\bigg|_{r_{0}}=0\,,\quad \mathcal{R}\big|_{r_{0}}=\frac{d\mathcal{R}}{dr}\bigg|_{r_{0}}=0\,.
\end{equation}
Consequently, the photon sphere radius $r_{0}$ associated with the maximum of the effective potential for black hole in the spacetime \eqref{ansatz} with arbitrary dimensions is the smallest value of the roots of the following equation
\begin{equation}\label{r0}
r_{0}h'(r_{0})-2h(r_{0})=0\,,
\end{equation}
where a prime stands for radial derivative.

\subsection{Geometrical shapes of shadow}

In this section we aim to find the shadow shape and size of the black holes in the spacetime \eqref{ansatz} with arbitrary dimensions. To do this, we begin with the definition of two impact parameters $\xi$ and $\eta$. These impact parameters as functions of the constants of motion $E$, $L$, and $\mathcal{K}$ can characterize the properties of photons near black holes. They define as follows
\begin{equation}\label{xieta1}
\xi=\frac{L}{E}\,,\qquad\eta=\frac{\mathcal{K}}{E^{2}}\,.
\end{equation}
Therefore, one can rewrite the effective potential and also, the function $\mathcal{R}$ in terms of these impact parameters as
\begin{equation}\label{veffR}
V_{eff}=E^{2}\left\{\frac{f(r)}{r^{2}}\left(\eta+\xi^{2}\right)-\frac{f(r)}{h(r)}\right\}\,,\qquad
\mathcal{R}=E^{2}\left\{r^{4}\frac{f(r)}{h(r)}-r^{2}f(r)\left(\eta+\xi^{2}\right)\right\}\,.
\end{equation}
Finally, by inserting Eq. \eqref{veffR} into Eq. \eqref{photonra} one can find the following equation for two unknowns $\xi$ and $\eta$
\begin{equation}\label{xieta2}
\eta+\xi^{2}=\frac{r_{0}^{2}}{2f(r_{0})+r_{0}f'(r_{0})}\left\{4\left(\frac{f(r_{0})}{h(r_{0})}\right)+r_{0}
\left(\frac{f'(r_{0})h(r_{0})-f(r_{0})h'(r_{0})}{h^{2}(r_{0})}\right)\right\}\,.
\end{equation}
Therefore, the photon sphere radius achieved from Eq. \eqref{r0} yields the quantity $\eta+\xi^{2}$ using Eq. \eqref{xieta2}. One can see that $r_{0}$ has the dimension of the length and the quantity $\eta+\xi^{2}$ has the dimension of the length square.

\begin{figure}[htb]
\centering
\subfloat{\includegraphics[width=0.49\textwidth]{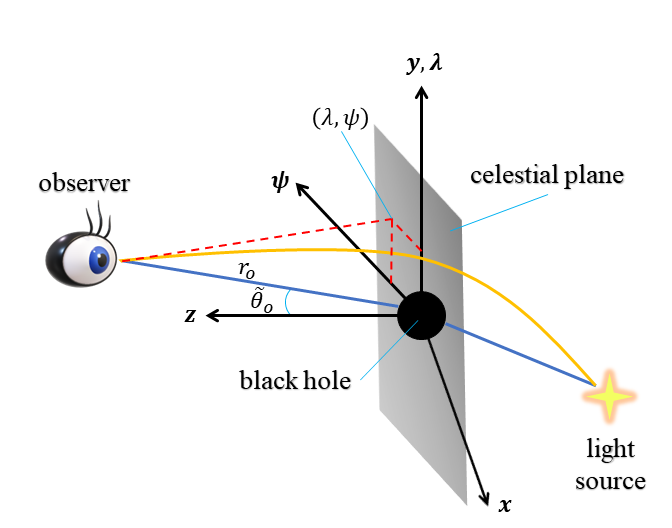}}
\caption{\label{Fig1}\small{\emph{The schematic of the celestial coordinates on the far observer's sky in which $r_{o}$ is the spatial separation between the far distant observer and the black hole, and $\tilde{\theta}_{o}$ is the angular coordinate of the far observer. Therefore, the location of the far observer is characterized with $(r_{o},\tilde{\theta}_{o})$. The coordinates $(\lambda,\psi)$ are the apparent perpendicular distance of the image as seen from the axis of symmetry, and from its projection on the equatorial plane, respectively.}}}
\end{figure}

The celestial coordinates $\lambda$ and $\psi$ \cite{Vazquez:2003zm} are employed to characterize the geometrical shape of the shadow as seen on the observer's frame. Fig. \ref{Fig1} is a schematic of the celestial coordinates used in this paper.
These coordinates can be read as follows
\begin{equation}\label{ab}
\lambda=\lim_{r_{o}\rightarrow\infty}\left(\frac{r_{o}^{2}P^{(\theta_{n-2})}}{P^{(t)}}\right)\,,\qquad
\psi=\lim_{r_{o}\rightarrow\infty}\left(\frac{r_{o}^{2}P^{(\theta_{i})}}{P^{(t)}}\right)\,,
\end{equation}
where $\left[P^{(t)},P^{(\theta_{n-2})},P^{(\theta_{i})}\right]$ are its vi-tetrad momentum elements and $r_{o}$ is the distance between the observer and the black hole. On the equatorial plane, one finds $\lambda=-\xi$ and $\psi=\pm\sqrt{\eta}$. Therefore, we can gain the following outcome
\begin{equation}\label{Rs}
R_{s}^{2}\equiv\eta+\xi^{2}=\lambda^{2}+\psi^{2}\,,
\end{equation}
in which $R_{s}$ is the shadow radius in celestial coordinates. For non-rotating (static) black holes, the geometrical shape of the shadow is circle with radius $R_{s}$.

\subsection{Energy emission rate}

Black holes can radiate through the phenomenon known as Hawking radiation. At very high energy, the absorption cross-section generally oscillates around a limiting constant $\sigma_{lim}$. For a very far distant observer, however, the absorption cross-section advances toward the black hole shadow \cite{Wei:2013kza,Belhaj:2020rdb}. One can prove that $\sigma_{lim}$ is approximately equal to the photon sphere area, which in arbitrary dimensions can be represented as follows \cite{Wei:2013kza,Li:2020drn,Decanini:2011xw}
\begin{equation}\label{sigmalim}
\sigma_{lim}\approx\frac{\pi^{\frac{n-2}{2}}}{\Gamma\left[\frac{n}{2}\right]}R_{s}^{n-2}\,.
\end{equation}
Thus, the complete form of the energy emission rate of higher-dimensional black holes can be read as
\begin{equation}\label{radirate}
\frac{d^{2}E(\varpi)}{d\varpi dt}=\frac{2\pi^{2}\sigma_{lim}}{e^{\frac{\varpi}{T}}-1}\varpi^{n-1}\,,
\end{equation}
where $\varpi$ is the emission frequency and $T$ is the Hawking temperature of the black hole.

\subsection{Deflection angle}

Here we aim to provide the framework for studying the deflection angle of higher-dimensional black holes in the spacetime \eqref{ansatz}. In this regard, we want to utilize the Gauss-Bonnet theorem \cite{Gibbons:2008rj,Arakida:2017hrm}. We first should find the optical metric on the equatorial hyperplane $\theta_{i}=\pi/2$ in the spacetime \eqref{ansatz}. Then, on this hyperplane, we set $d\theta_{n-2}^{2}=d\phi^{2}$ to find
\begin{equation}\label{optmet1}
ds^{2}=-h(r)dt^{2}+\frac{dr^{2}}{f(r)}+r^{2}d\phi^{2}\,.
\end{equation}
Then, for the considered null geodesics for which $ds^{2}=0$, the optical metric reads as follows
\begin{equation}\label{optmet2}
dt^{2}=\frac{dr^{2}}{h(r)f(r)}+\frac{r^{2}}{h(r)}d\phi^{2}\,.
\end{equation}
For this optical metric, we can calculate the Gaussian optical curvature $K=\frac{\bar{R}}{2}$ in which $\bar{R}$ is the Ricci scalar of the metric \eqref{optmet2} as follows
\begin{equation}\label{gauoptcur1}
K=\frac{2\,rh(r)f(r)h''(r)-2\,rf(r)h'(r)^{2}+h(r)h'(r)\left\{rf'(r)+2f(r)\right\}-2f'(r)h(r)^{2}}{2\,rh(r)}\,.
\end{equation}

In order to calculate the deflection angle, one should consider a non-singular manifold $\mathcal{D}_{\tilde{R}}$ with a geometrical size $\tilde{R}$ to employ the Gauss-Bonnet theorem, so that \cite{Gibbons:2008rj,Arakida:2017hrm}
\begin{equation}\label{gaubon1}
\int\int_{\mathcal{D}_{\tilde{R}}}KdS+\oint_{\partial\mathcal{D}_{\tilde{R}}}kdt+\sum_{i}\varphi_{i}=2\pi\zeta(\mathcal{D}_{\tilde{R}})\,,
\end{equation}
where $dS=\sqrt{\bar{g}}drd\phi$ and $dt$ are the surface and line element of the optical metric \eqref{optmet2}, respectively, $\bar{g}$ is the determinant of the optical metric, $k$ denotes the geodesic curvature of $\mathcal{D}_{\tilde{R}}$, and $\varphi_{i}$ is the jump (exterior) angle at the $i$-th vertex, and also, $\zeta(\mathcal{D}_{\tilde{R}})$ is the Euler characteristic number of $\mathcal{D}_{\tilde{R}}$. One can set $\zeta(\mathcal{D}_{\tilde{R}})=1$. Then, considering a smooth curve $y$, which has the tangent vector $\dot{y}$ and acceleration vector $\ddot{y}$, the geodesic curvature $k$ of $y$ can be defined as follows where the unit speed condition $\tilde{g}\left(\dot{y},\dot{y}\right)=1$ is employed
\begin{equation}\label{geocur}
k=\tilde{g}\left(\nabla_{\dot{y}\dot{y},\ddot{y}}\right)\,,
\end{equation}
which is a measure of deviations of $y$ from being a geodesic. In the limit $\tilde{R}\rightarrow\infty$, two jump angles $\varphi_{s}$ (of source) and $\varphi_{o}$ (of observer) will become $\pi/2$, i.e, $\varphi_{s}+\varphi_{o}\rightarrow\pi$. Considering $C_{\tilde{R}}:=r(\phi)$, we have $k(C_{\tilde{R}})=|\nabla_{\dot{C}_{\tilde{R}}}\dot{C}_{\tilde{R}}|\,\,\,\myeqq\,\,\, 1/\tilde{R}$ and therefore, we can find $\lim_{\tilde{R}\rightarrow\infty}dt=\tilde{R}d\phi$. Hence, $k(C_{\tilde{R}})dt=d\phi$. Consequently, the Gauss-Bonnet theorem will reduce to the following form
\begin{equation}\label{gaubon2}
\int\int_{\mathcal{D}_{\tilde{R}}}KdS+\oint_{C_{\tilde{R}}}kdt\,\,\,\myeq\,\,\,\int\int_{\mathcal{D}_{\infty}}KdS+\int_{0}^{\pi+\Theta}d\phi=\pi\,.
\end{equation}
Finally, using the straight light ray approximation $r(\phi)=\xi/\sin[\phi]$, the Gauss-Bonnet theorem results in the following expression to calculate the deflection angle (for more details, see Refs. \cite{Gibbons:2008rj,Arakida:2017hrm} and references therein)
\begin{equation}\label{defang}
\Theta=\pi-\int_{0}^{\pi+\Theta}d\phi=-\int_{0}^{\pi}\int_{\frac{\xi}{\sin[\phi]}}^{\infty}KdS\,.
\end{equation}

\subsection{Shadow observables}

Black hole shadow observables can provide strong evidence for the existence of black holes. These observables refer to the features of the shadow casted by a black hole on its surrounding bright accretion disk. They are obtained from the images of the event horizon of a black hole, which can be captured currently utilizing EHT. Studying black hole shadow observables can provide us with valuable information about the properties of black holes \cite{Kuang:2022ojj,Ghosh:2022jfi,Meng:2022kjs,Afrin:2021wlj}. The size and shape of the shadow can give us insights into the black hole parameters, which are expected to be constrained from the EHT data. Overall, studying black hole shadow observables is an important tool for understanding the mysterious and fascinating phenomena of black holes. To introduce the shadow observables, we propose that the observer is at the equatorial plane, where the angular coordinate of the observer or the inclination angle is $\tilde{\theta}_{o}=\pi/2$.

\begin{figure}[htb]
\centering
\subfloat{\includegraphics[width=0.35\textwidth]{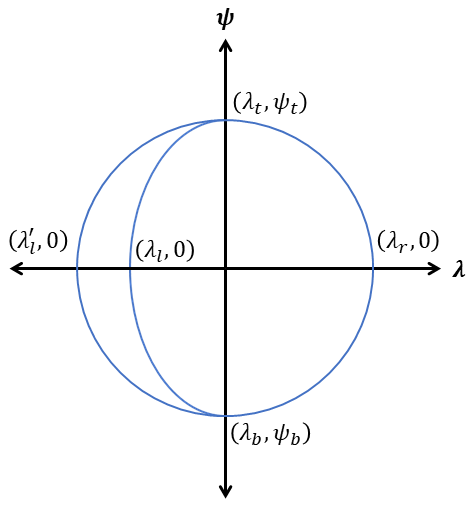}}
\caption{\label{Fig2}\small{\emph{Illustration of the shadow reference circle in the celestial coordinates.}}}
\end{figure}
Hioki and Maeda \cite{Hioki:2009na} suggested two characterized observables, $\tilde{R}_{s}$ and $\delta_{s}$ in order to investigate the size and distortion of black hole shadows. Based on Hioki-Maeda method, one can approximately describe the shadow of the black hole by a reference circle with the radius $\tilde{R}_{s}$ so that $\delta_{s}$ is the deviation of the left edge of the real shape of the shadow from the boundary of this reference circle \cite{Hioki:2009na}. In other words, $\tilde{R}_{s}$ is the shadow size and $\delta_{s}$ indicates the deformation of shadow shape from this circle of reference. The shadow reference circle in the celestial coordinates at the top, bottom, right, and left edges are located at $(\lambda_{t}, \psi_{t})$, $(\lambda_{b}, \psi_{b})$, $(\lambda_{r}, 0)$, and $(\lambda'_{l}, 0)$. Moreover, the leftmost edge of the shadow is located at $(\lambda_{l}, 0)$. We note that the indices $t$, $b$, $r$, and $l$ stand for the top, bottom, right, and left edge of the shadow. Figure \ref{Fig2} is the schematic of the shadow reference circle in the celestial coordinates. With these preliminaries, one can define these observables as
\begin{equation}\label{size}
\tilde{R}_{s}=\frac{\left(\lambda_{t}-\lambda_{r}\right)^{2}+\psi_{t}^{2}}{2\left|\lambda_{r}-\lambda_{t}\right|}\,,
\end{equation}
and
\begin{equation}\label{distostion}
\delta_{s}=\frac{\left|\lambda_{l}-\lambda'_{l}\right|}{\tilde{R}_{s}}\,.
\end{equation}

Kumar and Ghosh \cite{Kumar:2018ple} proposed that the shadow of some irregular black holes cannot be correctly characterized by $\tilde{R}_{s}$ and $\delta_{s}$ due to certain symmetry requirements in shadow shapes. Furthermore, due to noisy data, the shadow form may not be perfectly circular. Therefore, they introduced two new characterized observables, the shadow area $A_{s}$ and oblateness $D_{s}$ to describe haphazard shadows of any shape (not just circular shape), which are defined as follows
\begin{equation}\label{sharea}
A_{s}=2\int\psi(r_{0})\,d\lambda(r_{0})=2\int_{r_{0}^{-}}^{r_{0}^{+}}\left(\psi(r_{0})\frac{\lambda(r_{0})}{dr_{0}}\right)dr_{0}\,,
\end{equation}
and
\begin{equation}\label{oblateness}
D_{s}=\frac{\lambda_{r}-\lambda_{l}}{\psi_{t}-\psi_{b}}\,,
\end{equation}
where $r_{0}^{\pm}$ are retrograde and prograde orbits at the equatorial plane, respectively.

For non-rotating (spherically symmetric) black holes, as in the present study, one can verify that the shadow distortion can be eliminated so that $\delta_{s}=0$ and the shadow oblateness equals unity, i.e., $D_{s}=1$ \cite{Meng:2022kjs,Afrin:2021imp}. This indicates that for non-rotating black holes, the shadow shape is a perfect circle. Additionally, the retrograde and prograde orbits are not accessible for non-rotating black holes \cite{Meng:2022kjs,Afrin:2021imp}. In the subsequent section, however, we compare the shadow size of M87* supermassive black hole with the ALF and ALAdS charged higher-dimensional black holes in EHM gravity to constrain the electric charge and cosmological constant together with coupling constants of the EHM theory by following the procedure introduced by the EHT collaborations in Ref. \cite{EventHorizonTelescope:2021dqv}.

\section{Shadow and deflection angle of the higher-dimensional ALF and ALA\lowercase{d}S black holes in the EHM gravity}\label{shadef:alfads}

In this section, we aim to study the shadow and deflection angle of the higher-dimensional ALF and ALAdS black holes in the EHM gravity utilizing the general framework expressed in the previous section. To do this end, we apply the line element of these black holes in EHM gravity on the formulas deduced in the framework to investigate how dimensionality, electric charge, and cosmological constant in EHM gravity affect the shadow and deflection angle behavior. In this regard, we will see whether the shadow behavior of black holes are dependent to dimensionality, electric charge and cosmological constant as the spacetime features in addition to the black hole parameters.

\subsection{ALF charged black hole with extra dimensions}

To study the shadow and deflection angle of the ALF black hole in EHM gravity with extra dimensions, we arbitrarily consider the electric charge values as $Q=0.1, 0.5, 1, 1.5,$ and $2$. Also, we take into account that the extra dimensions count $n=4,5,\ldots,11$ (note that $n=4$ stands for one temporal in addition to three spatial dimensions as usual).

\subsubsection{Effective potential}

First, we want to check the behavior of the effective potential for the ALF black hole with extra dimensions in the EHM gravity. Inserting Eqs. \eqref{fralf} and \eqref{hralf} into Eq. \eqref{veff} results in the effective potential for the higher-dimensional ALF charged black hole as follows
\begin{equation}\label{effalf}
\begin{split}
V_{eff} & =\frac{16(n-2)^{2}(n-3)^{2}r^{4n-2}}{\left(q^{2}r^{6}-4(n-2)(n-3)r^{2n}\right)^{2}}\\
& \times\bigg\{\left(\mathcal{K}+L^{2}\right)\left(1-\frac{\mu}{r^{n-3}}+\frac{q^{2}}{2(n-2)(n-3)r^{2(n-3)}}-\frac{q^{4}}{48(n-2)^{2}(n-3)^{2}r^{4(n-3)}}\right)
-E^{2}r^{2}\bigg\}\,.
\end{split}
\end{equation}

\begin{figure}[htb]
\centering
\subfloat[\label{Fig3a} for $Q=0.5$]{\includegraphics[width=0.49\textwidth]{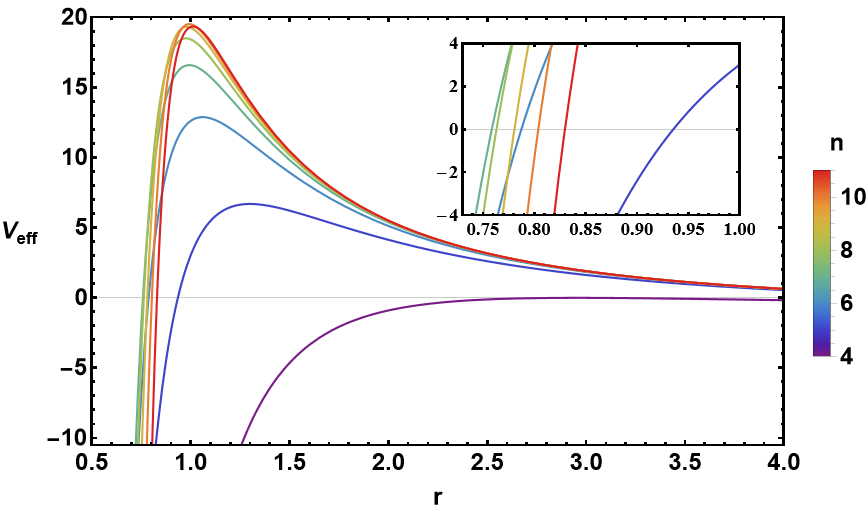}} \\
\subfloat[\label{Fig3b} for $n=4$]{\includegraphics[width=0.49\textwidth]{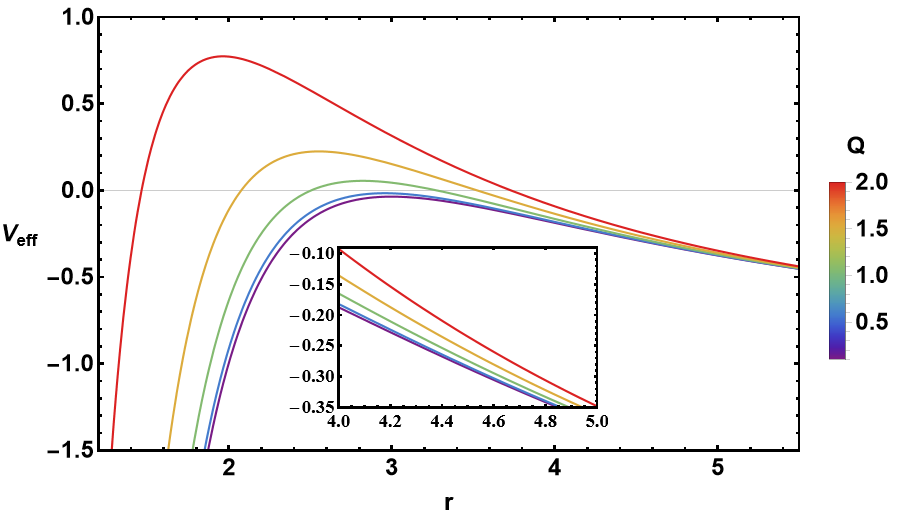}}
\,\,\,
\subfloat[\label{Fig3c} for $n=5$]{\includegraphics[width=0.46\textwidth]{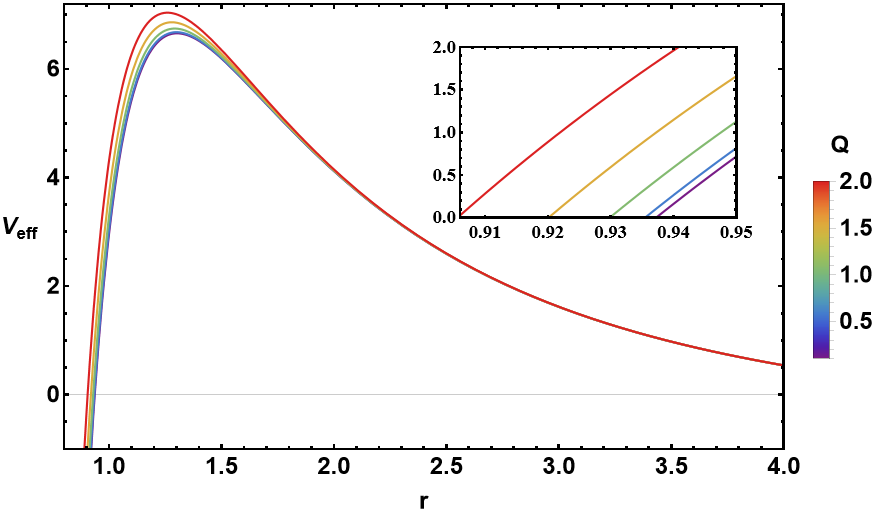}}
\caption{\label{Fig3}\small{\emph{The graph of the radial evolution of the effective potential for the ALF charged black hole with extra dimensions in EHM gravity for different values of $n$ and $Q$ in which we set $M=1$.}}}
\end{figure}
Fig. \ref{Fig3} depicts the behavior of the effective potential for the ALF charged black hole with extra dimensions as a function of radial coordinate $r$ for different values of $n$ and $Q$. In this figure, the effective potential peaks at the photon sphere radius $r_{0}$ associated with each value of $n$ and $Q$ and in the limit $r\rightarrow\infty$, the effective potential approaches a constant value. From Fig. \ref{Fig3a}, we see that for a fixed value of $Q$, the effective potential for the higher-dimensional ALF charged black hole increases by growing $n$. Also, we find from Fig. \ref{Fig3b} that increasing the value of $Q$ for $n=4$ leads to amplification of the effective potential for the black hole. However, form Fig. \ref{Fig3c} we find that for $n\geq 5$, although this amplifying of the effective potential continues but each curve of the effective potential corresponding to different values of $Q$ finally coincide. This fact shows that the impact of higher dimensions dominates the effect of the electric charge in the ALF black hole with higher dimensions. Since the location of the maximum of the effective potential for the black hole, i.e., the photon sphere radius $r_{0}$, characterizes the shadow boundary of the black hole, Fig. \ref{Fig3} shows that how $n$ and $Q$ affect the shadow boundary of the ALF black hole in the EHM gravity with extra dimensions.

\subsubsection{Geometrical shapes of shadow}

Now we are going to illustrate the geometrical shape of shadow of the ALF charged black hole with extra dimensions on the observer's sky in the celestial coordinates introduced in previous section. In this regard, we first should collect some numerical data for $r_{*}, r_{eh}, r_{0}$, and $\sqrt{\eta+\xi^{2}}$ associated with the black hole. Inserting Eq. \eqref{hralf} into Eq. \eqref{r0} yields the photon sphere radius for the higher-dimensional ALF black hole in the EHM gravity. Moreover, applying Eqs. \eqref{fralf} and \eqref{hralf} into Eq. \eqref{xieta2} and using Eq. \eqref{Rs} leads to find the radius of shadow circles for the higher-dimensional ALF charged black hole in celestial coordinates. In Table \ref{Table1} we collect the numerical data associated with $r_{*}$, $r_{eh}$, and $r_{0}$ for $n=4,5,\cdots,11$ and some different values of $Q$.
\begin{table}[htb]
        \centering
        \caption{Values of $r_{*}$, $r_{eh}$, and $r_{0}$ of the higher-dimensional ALF charged black hole for different values of $Q$ and $n$.}
        \label{Table1}
        \begin{tabular}{|c||c|c|c||c|c|c||c|c|c||c|c|c||c|c|c|}
        \hline
             & \multicolumn{3}{|c||}{$Q=0.1$} & \multicolumn{3}{|c||}{$Q=0.5$} & \multicolumn{3}{|c||}{$Q=1$} & \multicolumn{3}{|c||}{$Q=1.5$} & \multicolumn{3}{|c|}{$Q=2$}\\
             \cline{2-16}
          $n$   & $r_{*}$ & $r_{eh}$ & $r_{0}$ & $r_{*}$ & $r_{eh}$ & $r_{0}$ & $r_{*}$ & $r_{eh}$ & $r_{0}$ & $r_{*}$ & $r_{eh}$ & $r_{0}$ & $r_{*}$ & $r_{eh}$ & $r_{0}$ \\
            \hline\hline
            \multicolumn{1}{|c||}{$n=4$} & 0.03 & 1.99 & 2.99 & 0.17 & 1.96 & 2.95 & 0.35 & 1.86 & 2.82 & 0.53 & 1.66 & 2.56 & 0.70 & 1.23 & 2.05 \\
            \hline
            \multicolumn{1}{|c||}{$n=5$} & 0.08 & 0.92 & 1.302 & 0.19 & 0.919 & 1.301 & 0.27 & 0.91 & 1.29 & 0.33 & 0.90 & 1.28 & 0.38 & 0.89 & 1.27 \\
            \hline
            \multicolumn{1}{|c||}{$n=6$} & 0.14 & 0.782 & 1.061 & 0.24 & 0.781 & 1.06 & 0.30 & 0.779 & 1.059 & 0.34 & 0.777 & 1.057 & 0.38 & 0.774 & 1.054 \\
            \hline
            \multicolumn{1}{|c||}{$n=7$} & 0.19 & 0.755 & 0.993 & 0.29 & 0.754 & 0.992 & 0.34 & 0.753 & 0.992 & 0.38 & 0.752 & 0.991 & 0.41 & 0.751 & 0.99 \\
            \hline
            \multicolumn{1}{|c||}{$n=8$} & 0.24 & 0.761 & 0.978 & 0.33 & 0.76 & 0.976 & 0.38 & 0.759 & 0.975 & 0.42 & 0.758 & 0.975 & 0.44 & 0.757 & 0.975 \\
            \hline
            \multicolumn{1}{|c||}{$n=9$} & 0.29 & 0.78 & 0.98 & 0.38 & 0.778 & 0.979 & 0.43 & 0.777 & 0.979 & 0.46 & 0.777 & 0.979 & 0.48 & 0.776 & 0.979 \\
            \hline
            \multicolumn{1}{|c||}{$n=10$} & 0.34 & 0.8011 & 0.993 & 0.42 & 0.801 & 0.993 & 0.47 & 0.80 & 0.992 & 0.50 & 0.80 & 0.992 & 0.52 & 0.80 & 0.992 \\
            \hline
            \multicolumn{1}{|c||}{$n=11$} & 0.38 & 0.828 & 1.0114 & 0.47 & 0.827 & 1.0114 & 0.51 & 0.826 & 1.0113 & 0.54 & 0.826 & 1.0113 & 0.56 & 0.826 & 1.0112 \\
            \hline
        \end{tabular}
\end{table}

\begin{figure}[H]
\centering
\subfloat[\label{Fig4a} for $Q=0.1$]{\includegraphics[width=0.3\textwidth]{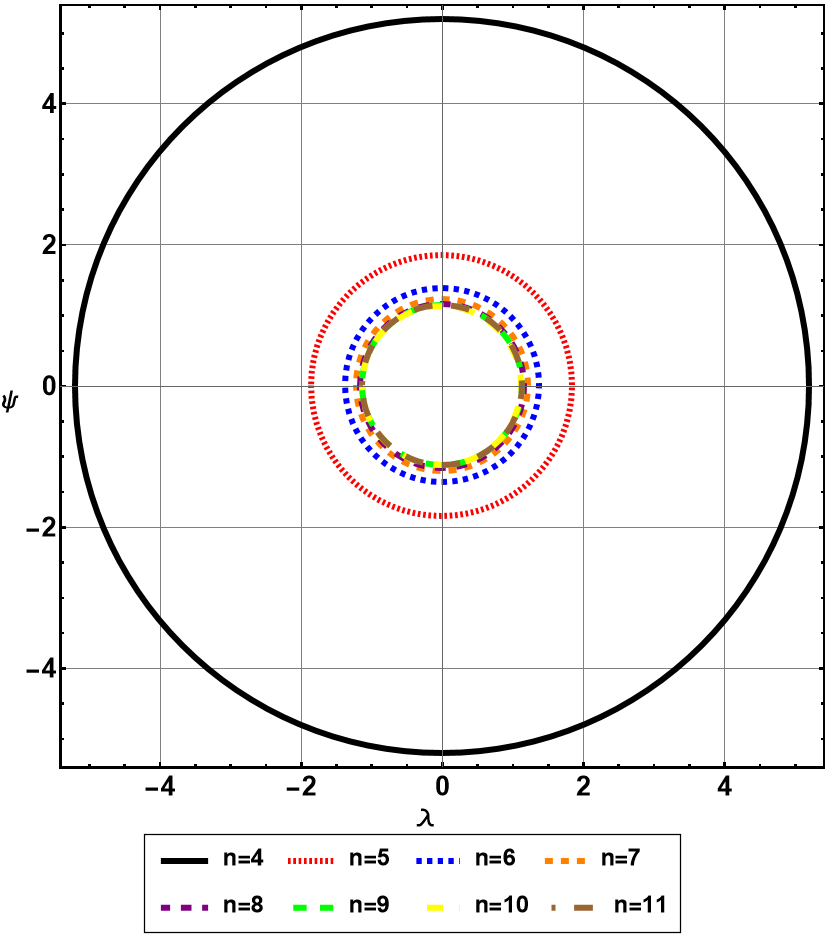}}
\,\,\,
\subfloat[\label{Fig4b} for $Q=0.5$]{\includegraphics[width=0.3\textwidth]{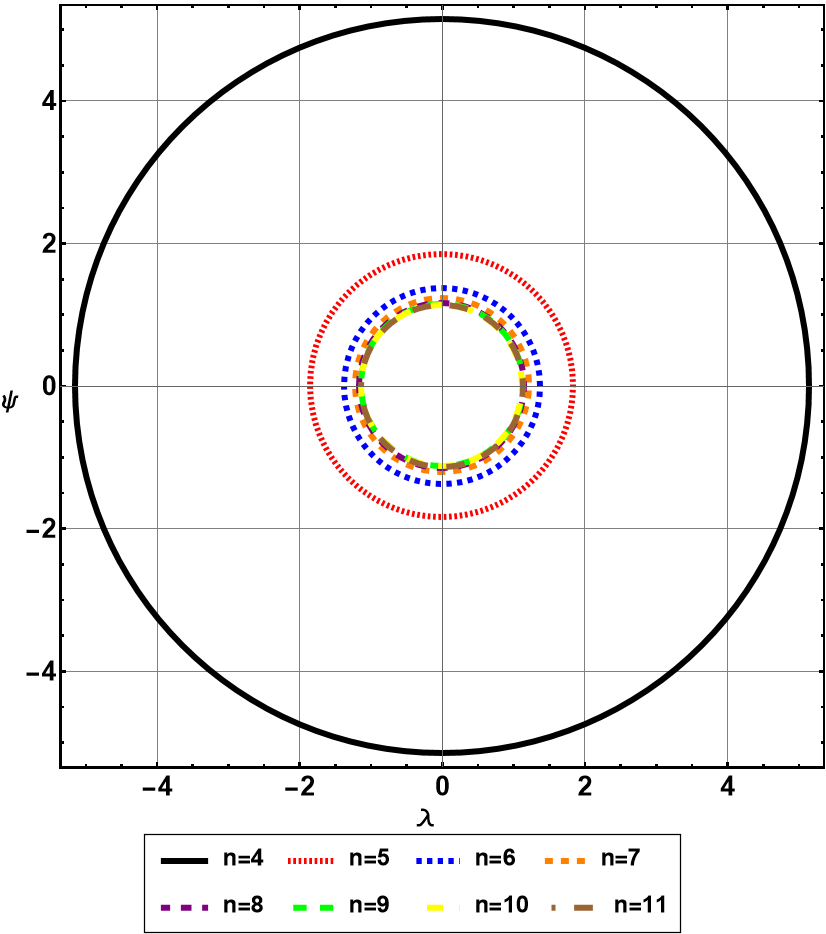}}
\,\,\,
\subfloat[\label{Fig4c} for $Q=1$]{\includegraphics[width=0.3\textwidth]{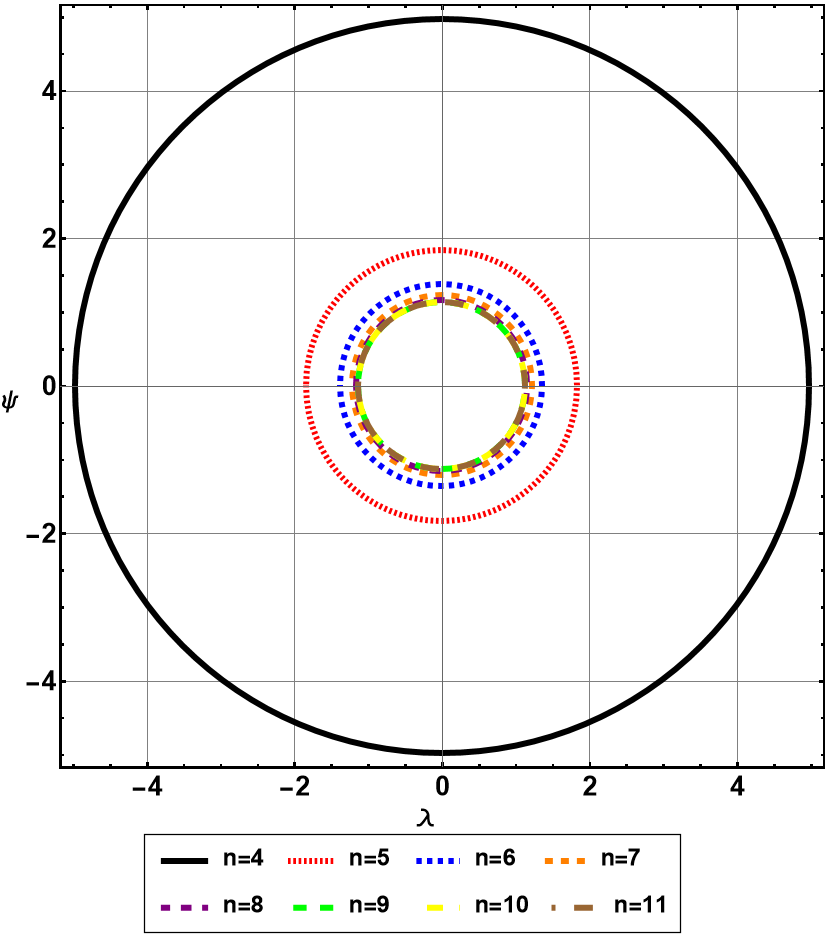}}
\\
\subfloat[\label{Fig4d} for $Q=1.5$]{\includegraphics[width=0.3\textwidth]{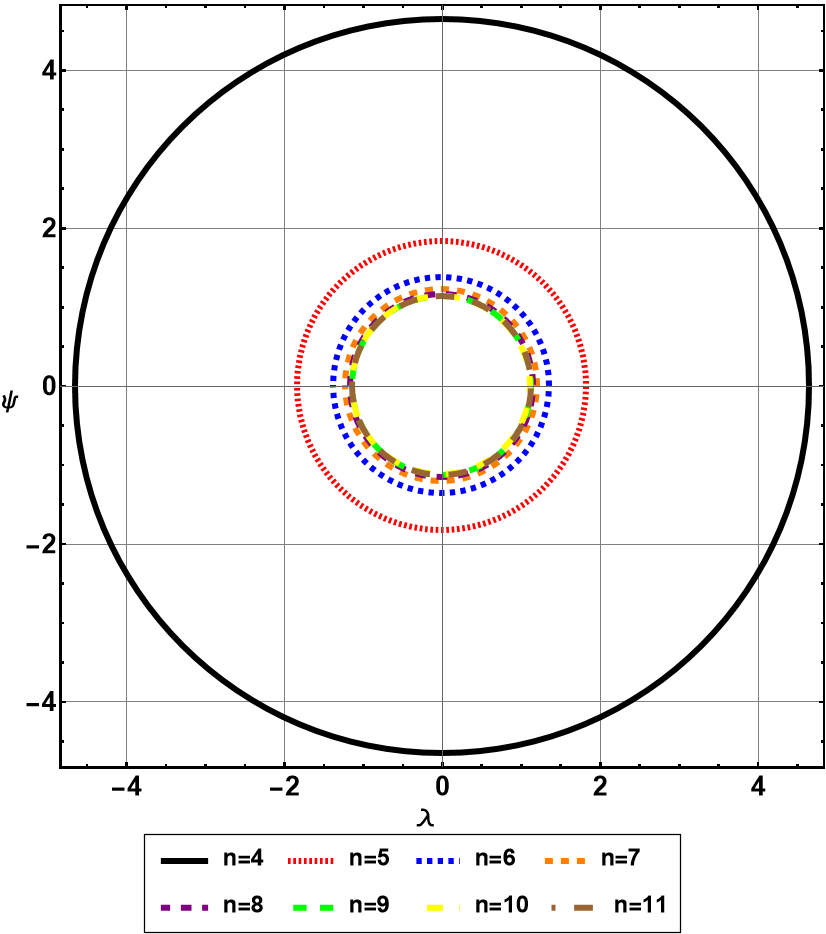}}
\,\,\,
\subfloat[\label{Fig4e} for $Q=2$]{\includegraphics[width=0.3\textwidth]{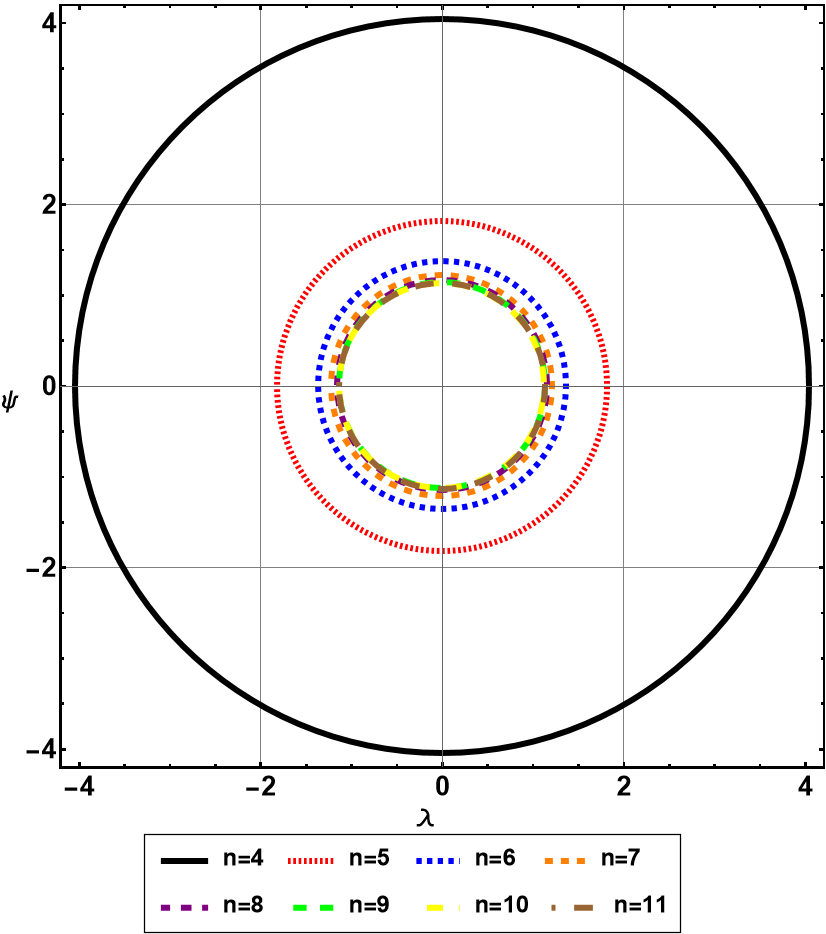}}
\caption{\label{Fig4}\small{\emph{Geometrical shape of the shadow of the higher-dimensional ALF charged black hole in celestial plane with $M=1$.}}}
\end{figure}
Based on the data provided in Table \ref{Table1}, one can plot the shadow circles of the ALF charged black hole with extra dimensions for different values of $Q$ and $n$. In Fig. \ref{Fig4}, we can see the geometrical shapes of shadow of the ALF charged black hole with extra dimensions in the celestial coordinates for different values of $Q$ and $n$. Each plot in Fig. \ref{Fig4} is for a fixed value of $Q$. From Fig. \ref{Fig4} we see that for a fixed value of $Q$, the shadow shapes of the black hole decrease with increasing dimension. Therefore, the extra dimensions affect the shadow of the black hole significantly by reducing the size of its geometrical shape.

\begin{figure}[htb]
\centering
\subfloat[\label{Fig5a} for $n=4$]{\includegraphics[width=0.295\textwidth]{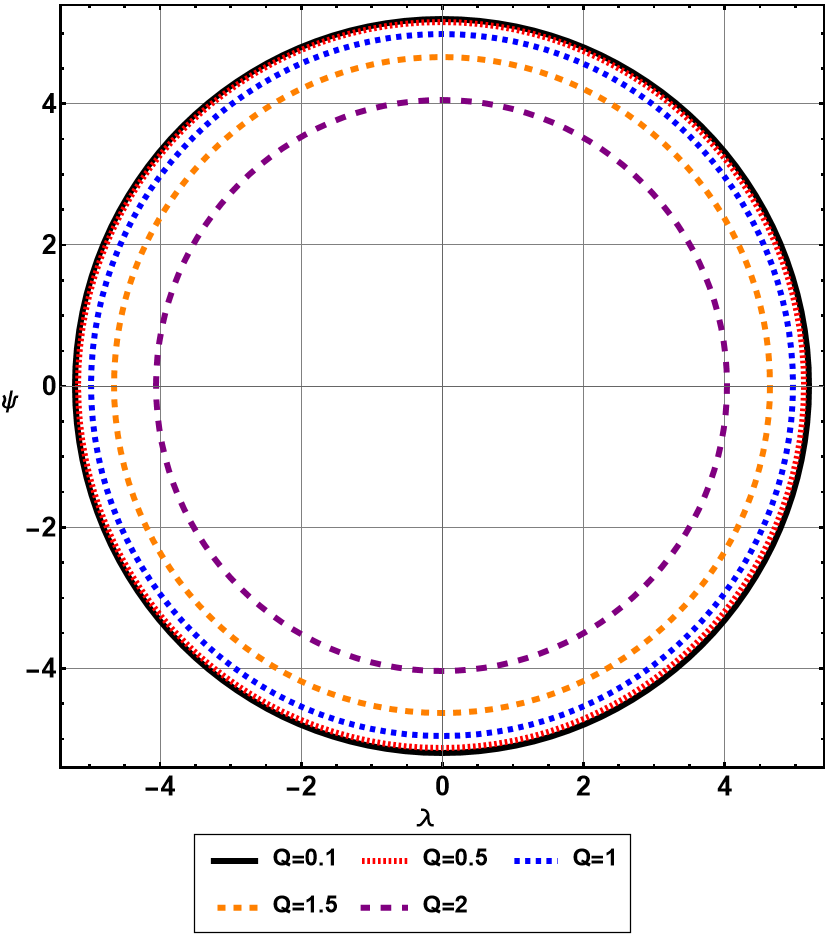}}
\,\,\,
\subfloat[\label{Fig5b} for $n=5$]{\includegraphics[width=0.3\textwidth]{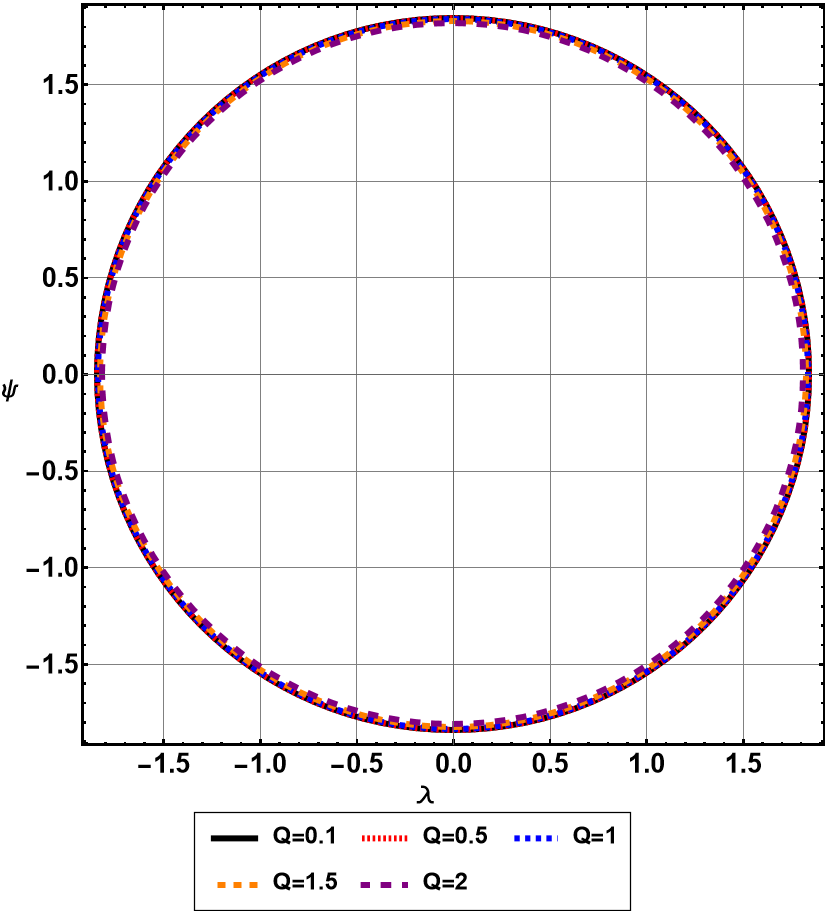}}
\caption{\label{Fig5}\small{\emph{Geometrical shape of the shadow of ALF charged black hole with higher dimensions in celestial plane with $M=1$.}}}
\end{figure}

Also, in Fig. \ref{Fig5}, we see the shadow circles of the ALF charged black hole with extra dimensions in the celestial coordinates for different values of $Q$ with $n=4$ in Fig. \ref{Fig5a} and $n=5$ in Fig. \ref{Fig5b}. In Fig. \ref{Fig5a} for $n=4$ we see that by increasing the electric charge value, the size of shadow circles decrease. In Fig. \ref{Fig5b}, however, for $n=5$ the shadow size of the ALF charged black hole with higher dimensions for each value of electric charge approach each other while they experience a reduction in their size in comparison with corresponding ones for $n=4$. One can see also the behavior for $n>5$. Therefore, Fig. \ref{Fig5} shows us that for $n\geq 5$, the shadow circles associated with different values of $Q$ coincide on each other. Therefore, the impact of the electric charge on the shadow of the black hole in EHM gravity is suppressible.

\subsubsection{Energy emission rate}

Now we apply the values of $r_{eh}$ from Table \ref{Table1} into Eq. \eqref{Htemalf} to gain the values of the Hawking temperature of the ALF charged black hole with extra dimensions for different values of $Q$ and $n$. Then, one can insert the values of $\sqrt{\eta+\xi^{2}}$ into Eq. \eqref{sigmalim} to obtain the values of $\sigma_{lim}$ for different values $Q$ and $n$ corresponding to the ALF charged black hole with higher dimensions. Finally, one can apply the values of the Hawking temperature and $\sigma_{lim}$ on Eq. \eqref{radirate} to gather the energy emission rate in terms of different values of $Q$ and $n$ associated with the higher-dimensional ALF black hole in EHM gravity.

\begin{figure}[htb]
\centering
\subfloat[\label{Fig6a} for $Q=0.5$ and $4\leq n\leq 7$]{\includegraphics[width=0.48\textwidth]{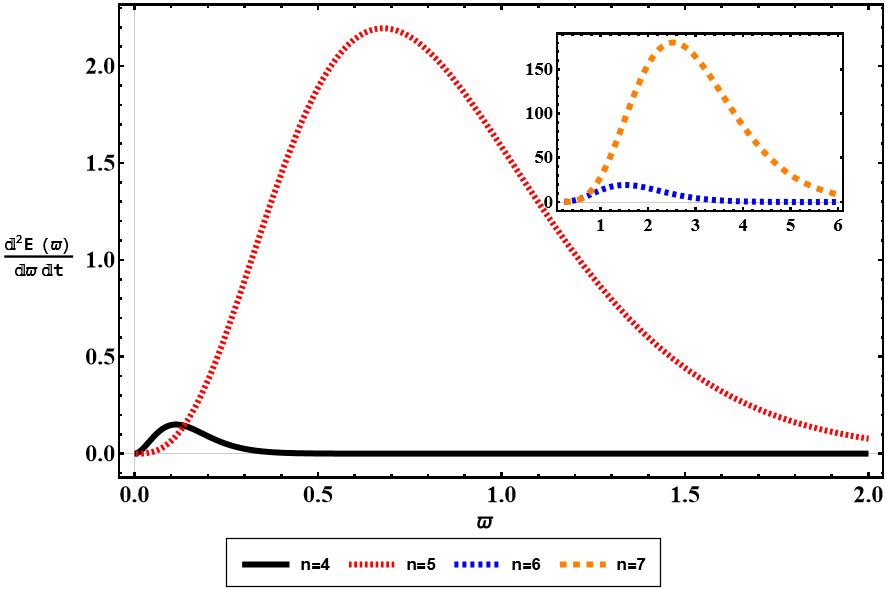}}
\,\,\,
\subfloat[\label{Fig6b} for $Q=0.5$ and $8\leq n\leq 11$]{\includegraphics[width=0.499\textwidth]{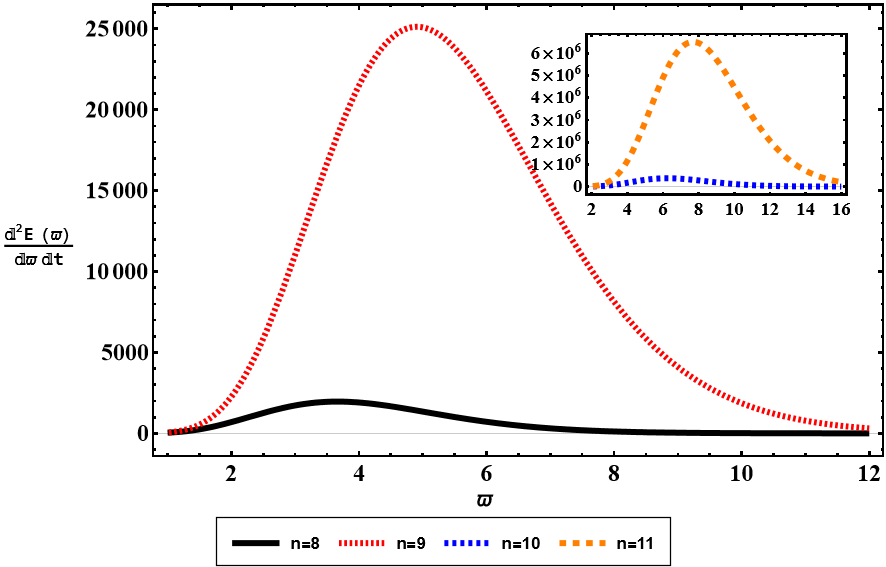}} \\
\subfloat[\label{Fig6c} for $n=4$]{\includegraphics[width=0.49\textwidth]{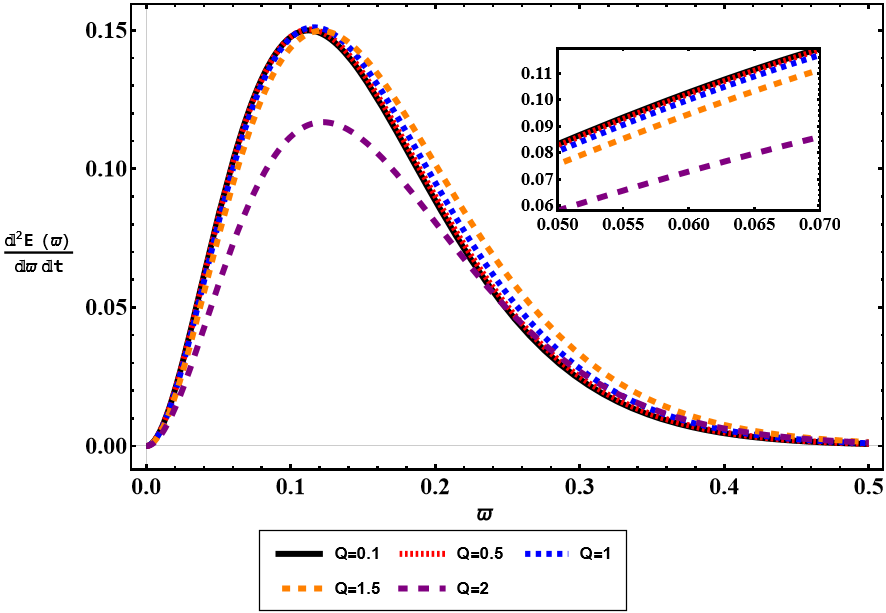}}
\,\,\,
\subfloat[\label{Fig6d} for $n=5$]{\includegraphics[width=0.49\textwidth]{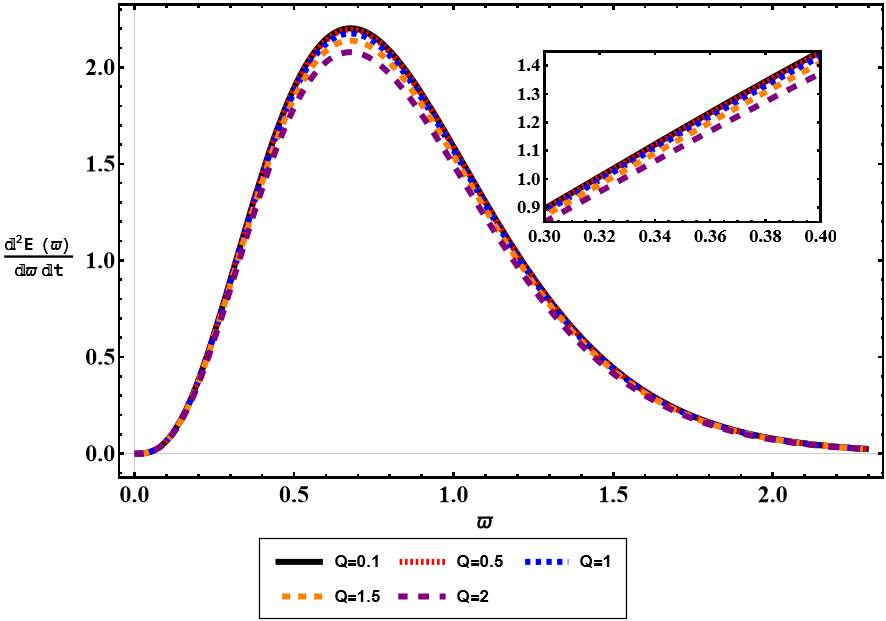}}
\caption{\label{Fig6}\small{\emph{The energy emission rate as a function of $\varpi$ for the higher-dimensional charged ALF black hole in EHM gravity for different values of $n$ with $Q$.}}}
\end{figure}
The energy emission rate for the higher-dimensional charged ALF black hole in EHM gravity is illustrated in Fig. \ref{Fig6} as a function of the emission frequency $\varpi$. From Figs. \ref{Fig6a} and \ref{Fig6b}, we see that for a fixed value of $Q$, the energy emission rate of the ALF charged black hole with extra dimensions extremely increases by growing $n$. Therefore, we found out that extra dimensions accelerate the evaporation of the ALF black hole in EHM gravity with higher dimensions. Additionally, from Fig. \ref{Fig6c} for $n=4$ one can see that increasing the electric charge value results in reducing the energy emission rate, especially for $Q=2$. From Fig. \ref{Fig6d}, however, one can find that for $n=5$ the energy emission rates of the ALF charged black hole with extra dimensions associated with each value of the electric charge approach each other while they experience an amplification in their values in comparison with corresponding ones for $n=4$ in Fig. \ref{Fig6c}. One can also verify such a behavior for $n>5$. Therefore, Figs. \ref{Fig6c} and \ref{Fig6d} show us that for $n\geq 5$, the energy emission rates associated with different values of $Q$ coincide with each other. So, although the impact of the electric charge is to amplify the energy emission rate of the black hole in EHM gravity, which causes black hole evaporation to accelerate, its effect is dominated by the impact of extra dimensions.

\subsubsection{Deflection angle}

Inserting Eqs. \eqref{fralf} and \eqref{hralf} into Eqs. \eqref{optmet2} and \eqref{gauoptcur1}, results in the Gaussian optical curvature for the higher-dimensional ALF black hole in EHM gravity, which up to the first order in mass and second order in electric charge of the black hole can be approximately found as follows
\begin{equation}\label{gauoptcur2}
K\approx-4\,\Gamma\left[\frac{n-1}{2}\right]\left\{\frac{M\pi^{\frac{3-n}{2}}(n-2)(n-3)^{2}r^{1-n}-2Q^{2}\pi^{3-n}r^{4-2n}\Gamma\left[\frac{n-1}{2}\right]}{(n-2)(n-3)}\right\}\,.
\end{equation}
Furthermore, the surface element of the optical metric \eqref{optmet2} for the higher-dimensional ALF black hole in EHM gravity corresponding to the metric coefficients \eqref{fralf} and \eqref{hralf} can be approximately found as follows
\begin{equation}\label{surele1}
dS=\sqrt{\bar{g}}\,drd\phi=\frac{r}{h(r)\sqrt{f(r)}}dtd\phi\approx rdrd\phi\,.
\end{equation}
Now, employing Eqs. \eqref{gauoptcur2} and \eqref{surele1} in the deflection angle formula \eqref{defang} leads to the deflection angle of the higher-dimensional ALF black hole in EHM gravity as follows
\begin{equation}\label{defang1}
\begin{split}
\Theta & =-\int_{0}^{\pi}\int_{\frac{\xi}{\sin[\phi]}}^{\infty}KdS\\
& \approx-\int_{0}^{\pi}\int_{\frac{\xi}{\sin[\phi]}}^{\infty}\Gamma\left[\frac{n-1}{2}\right]\left\{\frac{M\pi^{\frac{3-n}{2}}(n-2)(n-3)^{2}r^{1-n}-2Q^{2}\pi^{3-n}r^{4-2n}
\Gamma\left[\frac{n-1}{2}\right]}{(n-2)(n-3)}\right\}rdrd\phi\\
& =\frac{1}{\pi^{n}\xi^{2n-3}}\left\{4M\xi^{n}\pi^{\frac{n+4}{2}}\Gamma\left[\frac{n-2}{2}\right]-\frac{Q^{2}\xi^{3}\pi^{\frac{7}{2}}\Gamma\left[\frac{2n-5}{2}\right]
\left(\Gamma\left[\frac{n-3}{2}\right]\right)^{2}}{\Gamma[n-1]}\right\}\,.
\end{split}
\end{equation}

\begin{figure}[htb]
\centering
\subfloat[\label{Fig7a} for $Q=0.5$]{\includegraphics[width=0.49\textwidth]{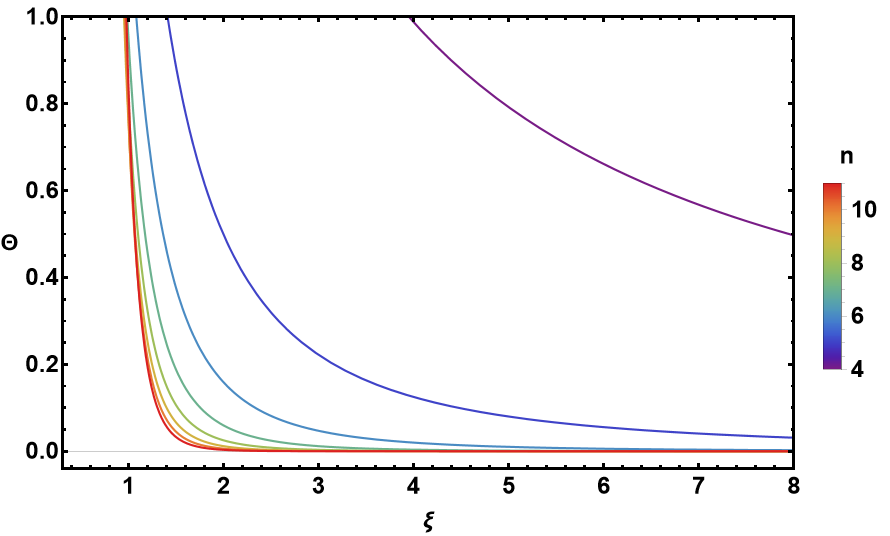}}\\
\subfloat[\label{Fig7b} for $n=4$]{\includegraphics[width=0.49\textwidth]{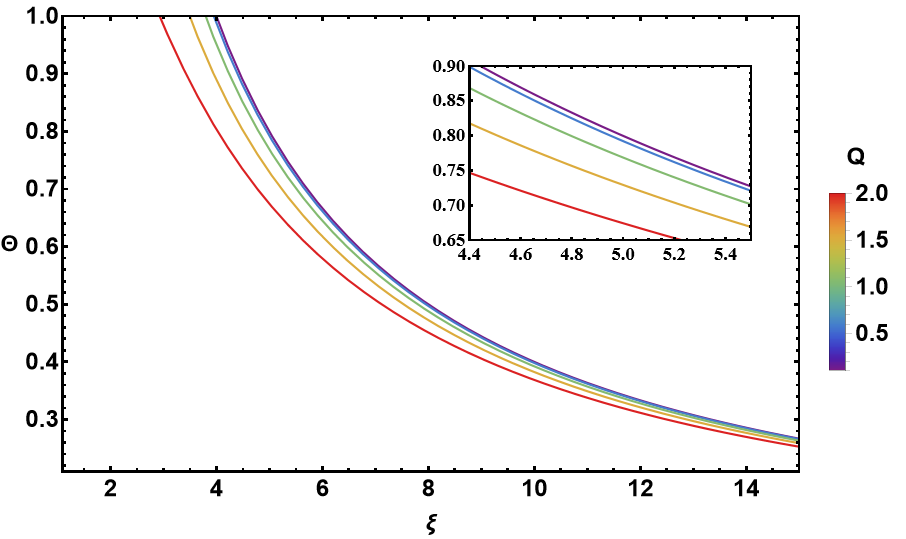}}
\,\,\,
\subfloat[\label{Fig7c} for $n=5$]{\includegraphics[width=0.49\textwidth]{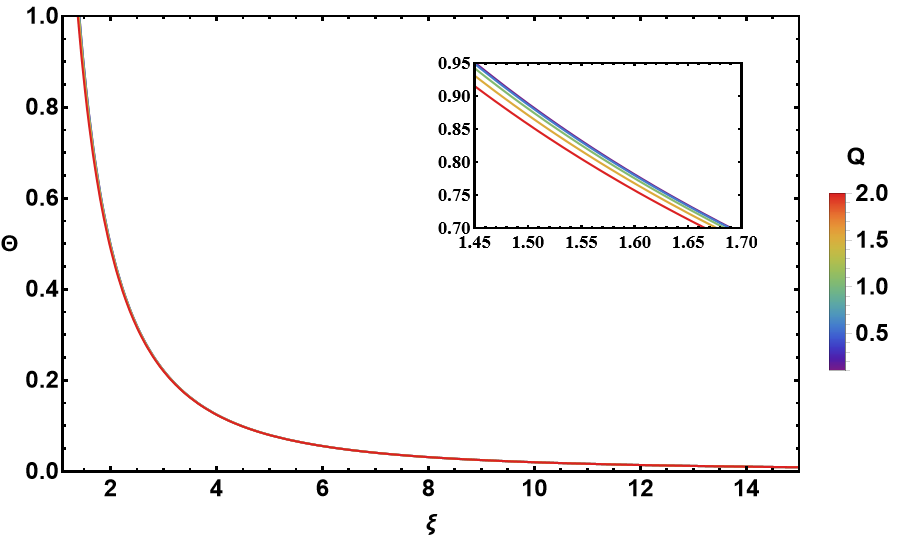}}
\caption{\label{Fig7}\small{\emph{The behavior of deflection angle of the higher-dimensional ALF charged black hole in EHM gravity in terms of $\xi$ for different values of $n$ and $Q$.}}}
\end{figure}
The behavior of the deflection angle of the higher-dimensional ALF charged black hole is illustrated in Fig. \ref{Fig7} for different values of $n$ with respect to $Q=0.5$ in Fig. \ref{Fig7a} and for different values of $Q$ with respect to $n=4, 5$ in Figs. \ref{Fig7b} and \ref{Fig7c}, respectively. From Fig. \ref{Fig7a}, we see that decreasing the value of the impact parameter $\xi$ results in extremely increasing the deflection angle of the black hole. Also, Fig. \ref{Fig7a} shows us that for a fixed value of electric charge, the deflection angle of the black hole reduces by growing the number of extra dimensions. In Fig. \ref{Fig7b} we see that for a fixed value of $n$, the deflection angle of the black hole decreases by increasing electric charge $Q$. However, as Fig. \ref{Fig7c} shows, for $n\geq 5$ the deflection angle curves of the black hole corresponding to each value of $Q$ coincide. This again shows that the effect of the electric charge is dominated by the impact of extra dimensions in the ALF charged black hole.

\subsubsection{Constraints from EHT observations of M87*}

Here we aim to compare the deduced shadow radius of the higher-dimensional ALF charged black hole in EHM gravity with the shadow size of supermassive black hole, M87* captured by EHT. Within $1$-$\sigma$ (68\%) confidence levels, one can find that the shadow size of M87* supermassive black hole captured by EHT lies within the interval \cite{EventHorizonTelescope:2021dqv}
\begin{equation}\label{M87}
4.31\leq R_{s, M87^{*}}\leq 6.08\,.
\end{equation}
Comparing this with the shadow size of the higher-dimensional ALF charged black hole in EHM gravity enables us to constrain the electric charge values.

\begin{figure}[htb]
\centering
\subfloat{\includegraphics[width=0.6\textwidth]{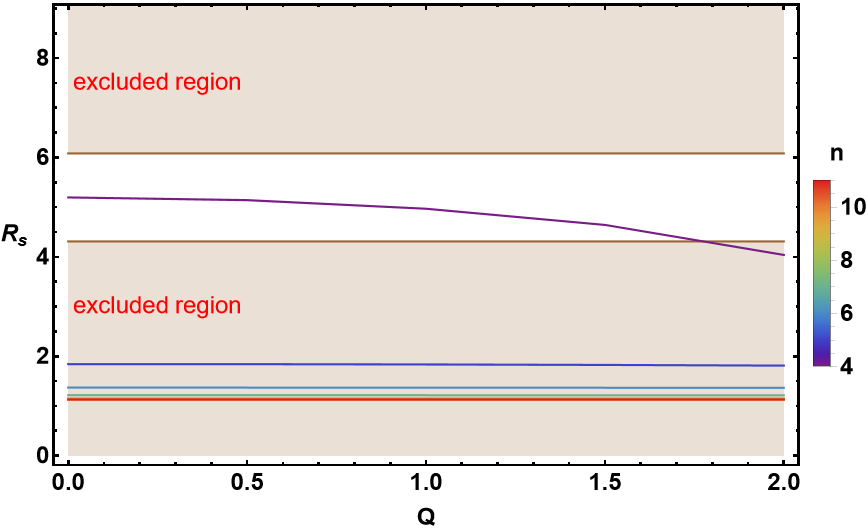}}
\caption{\label{Fig8}\small{\emph{The shadow radius of the higher-dimensional ALF charged black hole in EHM gravity in comparison with the shadow size of M87* captured via EHT within $1$-$\sigma$ confidence level versus the electric charge. The brown (shaded) area is the excluded region, which is inconsistent with the observations of EHT, while the white (unshaded) region is the $1$-$\sigma$ confidence level of EHT data.}}}
\end{figure}
Figure \ref{Fig8} indicates the behavior of the shadow radius of the higher-dimensional ALF charged black hole in EHM gravity in comparison with the EHT's shadow size of M87* within $1$-$\sigma$ uncertainties given in Eq. \eqref{M87} versus the electric charge. In Fig. \ref{Fig8}, the white (unshaded) region denotes the $1$-$\sigma$ confidence level while the brown (shaded) areas are the excluded regions, which are incompatible with the EHT observations associated with shadow radius of M87*. From Fig. \ref{Fig8}, we see that the shadows of the higher-dimensional ALF charged black hole in EHM gravity associated with $n=5,\ldots,11$ are incompatible with the observations of EHT. However, the shadow of the ALF charged black hole in EHM gravity with $n=4$ lies in the $1$-$\sigma$ confidence level, so that in the range $0\leq Q<1.8$ the shadow radius of the four dimensional black hole in EHM gravity has a good consistency with EHT observations. Moreover, like Table \ref{Table1} and Fig. \ref{Fig5a}, we see from Fig. \ref{Fig8} that decreasing the electric charge value leads to reduce the shadow radius of the four dimensional ALF charged black hole. The decreasing the electric charge, however, has no effect on the shadow of the higher-dimensional ALF charged black hole in EHM gravity associated with $n=5,\ldots,11$ since the impact of the electric charge in comparison with the extra dimension effect can be eliminated.

\subsection{ALA\lowercase{d}S charged black hole with extra dimensions}

Due to the complexity of the metric coefficients of the ALAdS charged black hole in EHM gravity with extra dimensions, we consider $n=4$ in addition to two odd dimensions, $n=5$ and $n=7$. Moreover, based on the previously mentioned condition $(\alpha+\gamma\Lambda)<0$ for the ALAdS case, we utilize two different sets $(\alpha=0.01,\,\gamma=0.51)$ and $(\alpha=0.015,\,\gamma=0.81)$ by considering five different values for the negative cosmological constant, $\Lambda=-0.02, -0.04, -0.06, -0.08$, and $-0.10$. Moreover, we set $Q=0.5$ to study the impact of extra dimensions and the negative cosmological constant.

\subsubsection{Effective potential}

As mentioned before, the effective potential plays a key role in studying shadow. We can find the effective potential for the ALAdS charged black hole with extra dimensions by inserting Eqs. \eqref{fralads}-\eqref{hrq} together with \eqref{hbarreven} for $n=4$, and \eqref{hbarrodd} for $n=5$ and $n=7$ into Eq. \eqref{veff}, which for $n=4$ yields
\begin{equation}\label{effadsnfour}
\begin{split}
V_{eff} & =\frac{4r^{4}(4+\beta\gamma)^{2}\left(3g^{2}r^{2}+1\right)^{2}}{\left(6g^{2}r^{4}(4+\beta\gamma)-q^{2}+8r^{2}\right)^{2}}\Bigg\{\frac{\left(\mathcal{K}+L^{2}\right)}
{12r^{6}(4+\beta\gamma)^{2}}\Bigg(-\sqrt{3}\,r^{3}\cot^{-1}\left[\sqrt{3}\,gr\right]\left(3g^{2}q^{2}-2\beta\gamma\right)^{2}g^{-1} \\
& +q^{4}\left(9g^{2}r^{2}-1\right)+48q^{2}r^{2}+12r^{3}(4+\beta\gamma)\left(r\left(-\beta\gamma+g^{2}r^{2}(4+\beta\gamma)+4\right)-\mu(4+\beta\gamma)\right)\Bigg)-E^{2}\Bigg\}
\end{split}
\end{equation}
and for $n=5$ and $n=7$ results in
\begin{equation}\label{effadson}
\begin{split}
V_{eff} & =\frac{(n-2)^{2}(4+\beta\gamma)^{2}\left(g^{2}r^{2}(n-1)+n-3\right)^{2}}{\left(g^{2}r^{2}(n-2)(n-1)(4+\beta\gamma)-q^{2}r^{6-2n}+4(n-3)(n-2)\right)^{2}}\\
& \times\Bigg\{\frac{\left(\mathcal{K}+L^{2}\right)}{r^{2}}\bigg(-\frac{\mu}{r^{n-3}}+\frac{8g^{2}r^{2}(2+\beta\gamma)+16}{(4+\beta\gamma)^{2}}
+\frac{2q^{2}}{(n-2)(n-3)(4+\beta\gamma)r^{2n-6}}\\
& -\frac{2\beta\gamma(n-3)q^{2}}{g^{2}(n-1)^{2}(n-2)(4+\beta\gamma)^{2}r^{2n-4}}
+\frac{(n-1)\beta^{2}\gamma^{2}g^{4}r^{4}}{(n-3)(4+\beta\gamma)^{2}}{}_{2}\mathrm{F}_{1}\left[1,\frac{n+1}{2};\frac{n+3}{2};\frac{(n-1)g^{2}r^{2}}{3-n}\right]\\
& +\frac{2\beta\gamma(n-3)^{2}q^{2}}{g^{4}(n+1)(n-1)^{2}(n-2)(4+\beta\gamma)^{2}r^{2n-2}}{}_{2}\mathrm{F}_{1}\left[1,\frac{n+1}{2};\frac{n+3}{2};\frac{3-n}{(n-1)g^{2}r^{2}}
\right]\\
& -\frac{q^{4}}{g^{2}(n-1)(n-2)^{2}(3n-7)(4+\beta\gamma)^{2}r^{2(2n-5)}}{}_{2}\mathrm{F}_{1}\left[1,\frac{3n-7}{2};\frac{3n-5}{2};\frac{3-n}{(n-1)g^{2}r^{2}}\right]\bigg)
-E^{2}\Bigg\}\,.
\end{split}
\end{equation}
\begin{figure}[htb]
\centering
\subfloat[\label{Fig9a} for $\Lambda=-0.02$]{\includegraphics[width=0.47\textwidth]{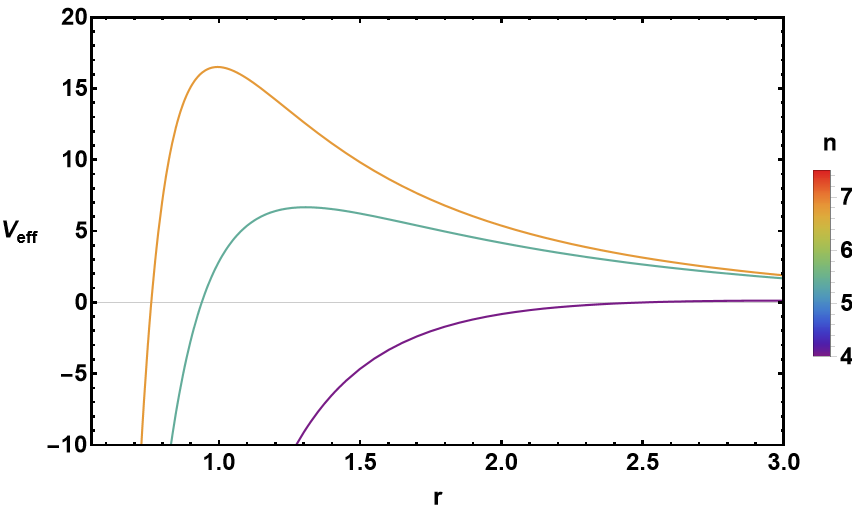}}
\,\,\,
\subfloat[\label{Fig9b} for $n=4$]{\includegraphics[width=0.49\textwidth]{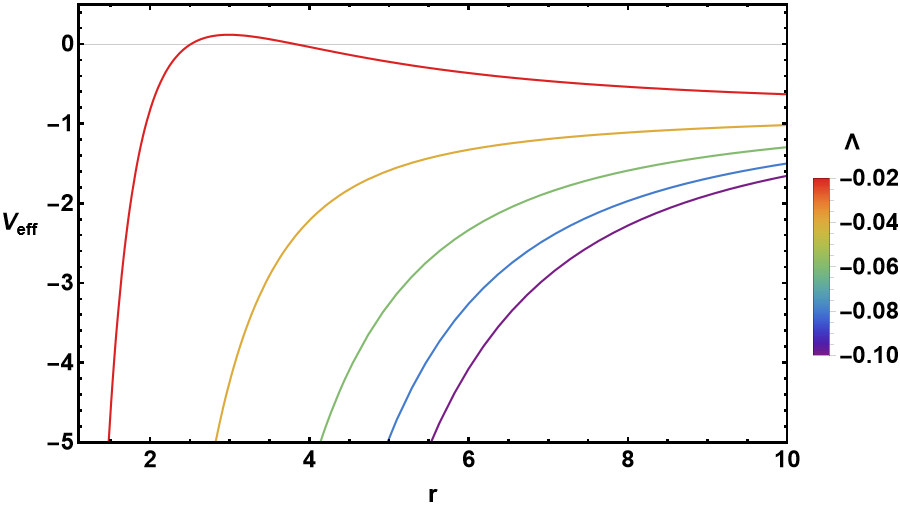}} \\
\subfloat[\label{Fig9c} for $n=5$]{\includegraphics[width=0.49\textwidth]{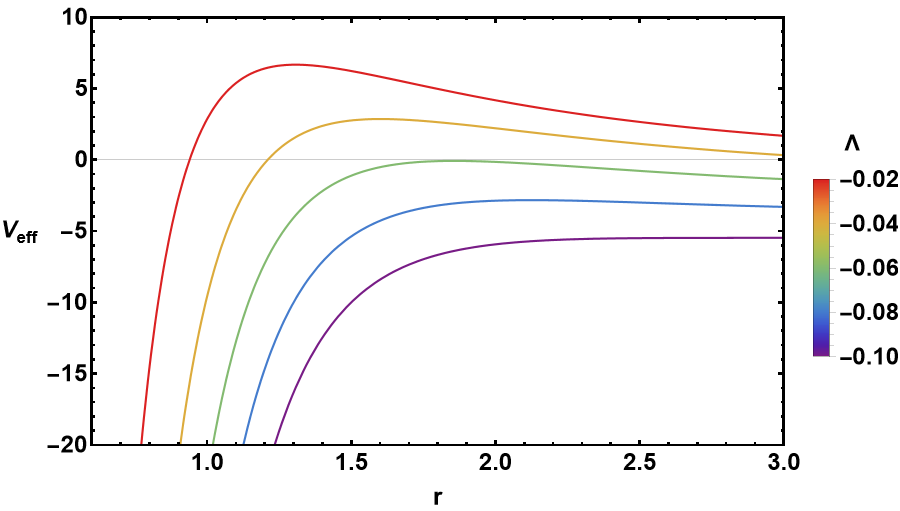}}
\,\,\,
\subfloat[\label{Fig9d} for $n=7$]{\includegraphics[width=0.49\textwidth]{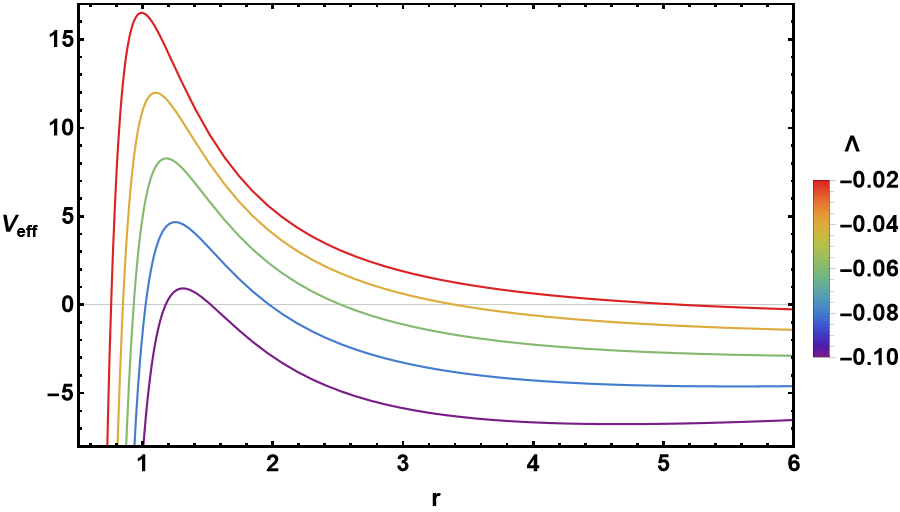}}
\caption{\label{Fig9}\small{\emph{The graph of radial evolution of the effective potential for the higher-dimensional ALAdS charged black hole in EHM gravity for different values of $n$ and $\Lambda$ in which we set $Q=0.5$ and $M=1$ using the set $(\alpha=0.01,\,\gamma=0.51)$.}}}
\end{figure}

Fig. \ref{Fig9} is the illustration of the effective potential versus $r$ for the ALAdS charged black hole with higher dimensions for different values of $n$ and $\Lambda$ for which we utilized the set $(\alpha=0.01,\,\gamma=0.51)$. We see from Fig. \ref{Fig9a} that for the fixed value of the cosmological constant $\Lambda=-0.02$, the effective potential for the higher-dimensional ALAdS black hole (like ALF one) in EHM gravity increases by growing $n$. Furthermore, Figs. \ref{Fig9b}, \ref{Fig9c}, and \ref{Fig9d} show that for a fixed value of $n$, increasing the value of $\Lambda$ (i.e., decreasing its absolute value) leads to amplify the effective potential for the higher-dimensional ALAdS charged black hole. This amplification becomes more remarkable by increasing the number of dimensions $n$ since in Fig. \ref{Fig9d} for $n=7$, the curve of effective potential related to $\Lambda=-0.02$ has much larger values than the corresponding ones in Figs. \ref{Fig9b} and \ref{Fig9c}. Consequently, form Fig. \ref{Fig9}, we find that the number of extra dimensions together with the cosmological constant have simultaneously an amplifying impact on the effective potential for the higher-dimensional ALAdS charged black hole in EHM gravity.

\subsubsection{Geometrical shapes of shadow}

One can characterize the geometrical shape of the shadow of the ALAdS charged black hole in EHM gravity with extra dimensions on the observer's frame utilizing the celestial coordinates. Applying Eqs. \eqref{hralads} and \eqref{hrq} together with \eqref{hbarreven} for $n=4$, and \eqref{hbarrodd} for $n=5$ and $n=7$ onto Eq. \eqref{r0} results in the radius of the photon sphere of the black hole. Also, one can gather the radius of shadow circles of the ALAdS black hole in EHM gravity with higher dimensions through inserting Eqs. \eqref{fralads}-\eqref{hrq} together with \eqref{hbarreven} for $n=4$, and \eqref{hbarrodd} for $n=5$ and $n=7$ into Eq. \eqref{xieta2} and making use of Eq. \eqref{Rs}. The numerical data associated with $r_{*}$, $r_{eh}$, and $r_{0}$ for different values of $\Lambda$ and $n$ utilizing the considered set $(\alpha=0.01,\,\gamma=0.51)$ are provided in Table \ref{Table2}.
\begin{table}[htb]
        \centering
        \caption{Numerical values of $r_{*}$, $r_{eh}$, and $r_{0}$ of the higher-dimensional ALAdS charged black hole in EHM gravity for different values of $\Lambda$ and $n$ considering the set $(\alpha=0.01,\,\gamma=0.51)$.}
        \label{Table2}
        \begin{tabular}{|c||c|c|c||c|c|c||c|c|c||c|c|c||c|c|c|}
        \hline
             & \multicolumn{3}{|c||}{$\Lambda=-0.02$} & \multicolumn{3}{|c||}{$\Lambda=-0.04$} & \multicolumn{3}{|c||}{$\Lambda=-0.06$} & \multicolumn{3}{|c||}{$\Lambda=-0.08$} & \multicolumn{3}{|c|}{$\Lambda=-0.10$}\\
             \cline{2-16}
          $n$   & $r_{*}$ & $r_{eh}$ & $r_{0}$ & $r_{*}$ & $r_{eh}$ & $r_{0}$ & $r_{*}$ & $r_{eh}$ & $r_{0}$ & $r_{*}$ & $r_{eh}$ & $r_{0}$ & $r_{*}$ & $r_{eh}$ & $r_{0}$ \\
            \hline\hline
            \multicolumn{1}{|c||}{$n=4$} & 0.1767 & 1.94 & 2.98 & 0.17669 & 4.66 & 9.43 & 0.17666 & 7.27 &  & 0.17664 & 8.79 &  & 0.17661 & 9.74 &  \\
            \hline
            \multicolumn{1}{|c||}{$n=5$} & 0.19367 & 0.92 & 1.307 & 0.19366 & 1.13 & 1.61 & 0.19365 & 1.301 & 1.86 & 0.19365 & 1.45 & 2.08 & 0.19364 & 1.58 & 2.28 \\
            \hline
            \multicolumn{1}{|c||}{$n=7$} & 0.290929 & 0.76 & 0.1 & 0.290926 & 0.84 & 1.1 & 0.290923 & 0.9 & 1.19 & 0.29092 & 0.95 & 1.25 & 0.290917 & 0.99 & 1.31 \\
            \hline
        \end{tabular}
\end{table}

\begin{figure}[H]
\centering
\subfloat[\label{Fig10a} for $\Lambda=-0.02$]{\includegraphics[width=0.29\textwidth]{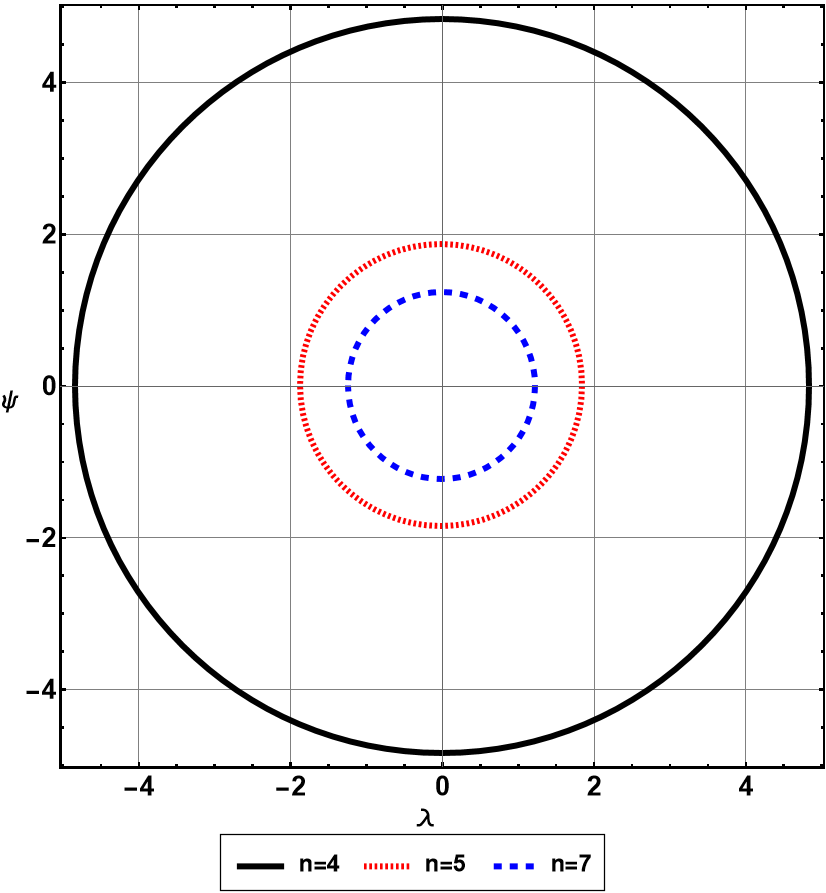}}
\,\,\,
\subfloat[\label{Fig10b} for $\Lambda=-0.04$]{\includegraphics[width=0.2905\textwidth]{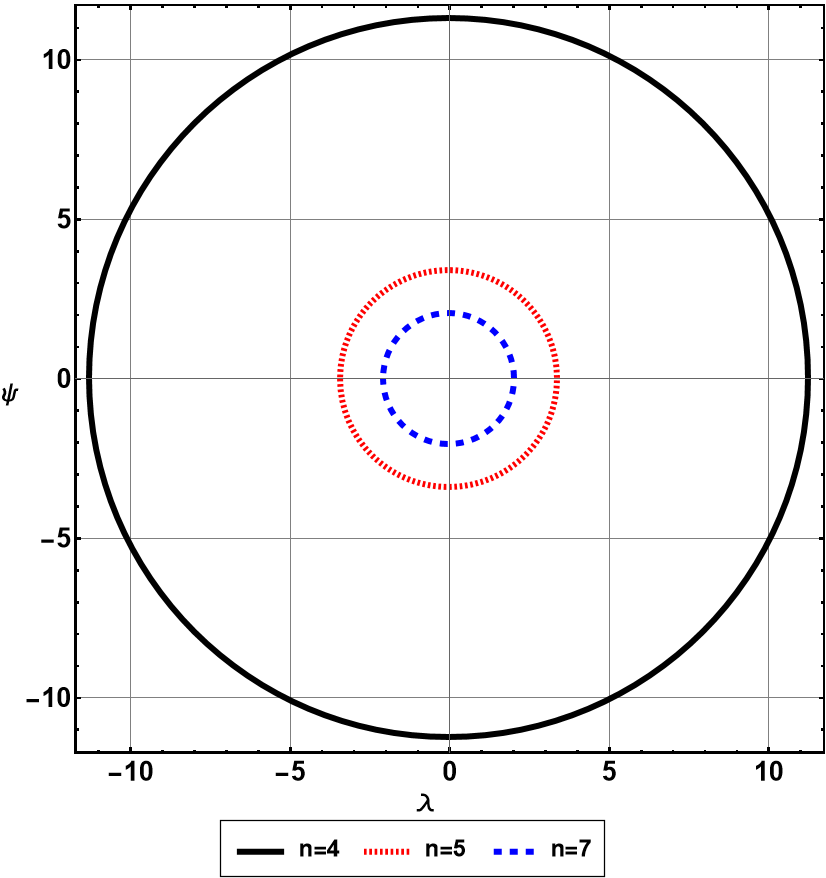}}
\,\,\,
\subfloat[\label{Fig10c} for $\Lambda=-0.06$]{\includegraphics[width=0.29\textwidth]{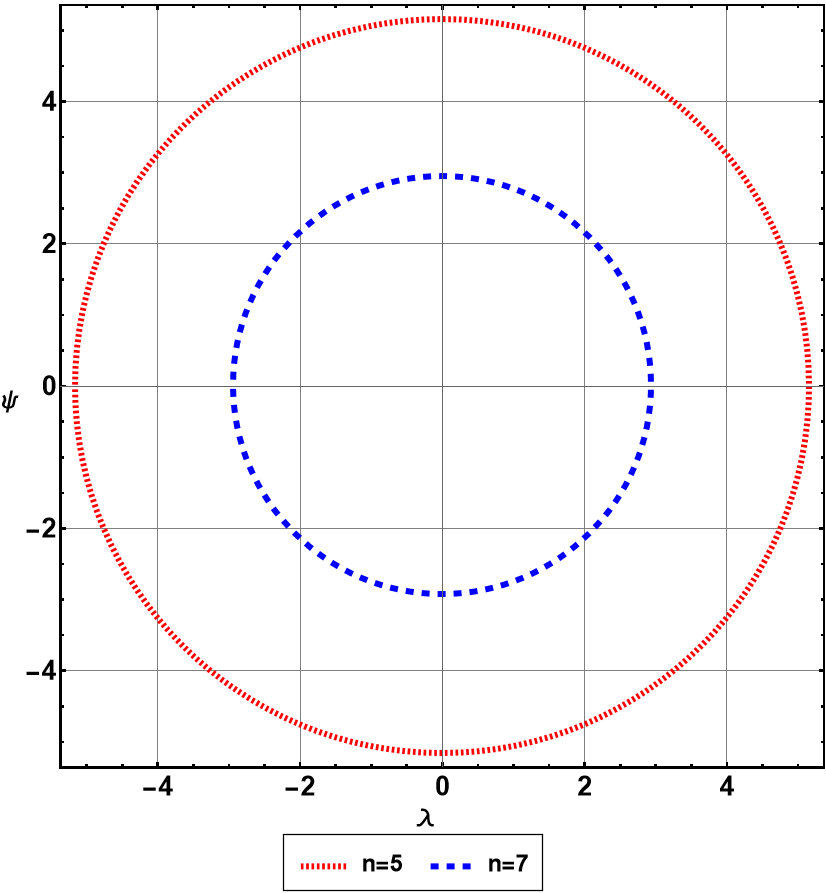}}
\\
\subfloat[\label{Fig10d} for $\Lambda=-0.08$]{\includegraphics[width=0.29\textwidth]{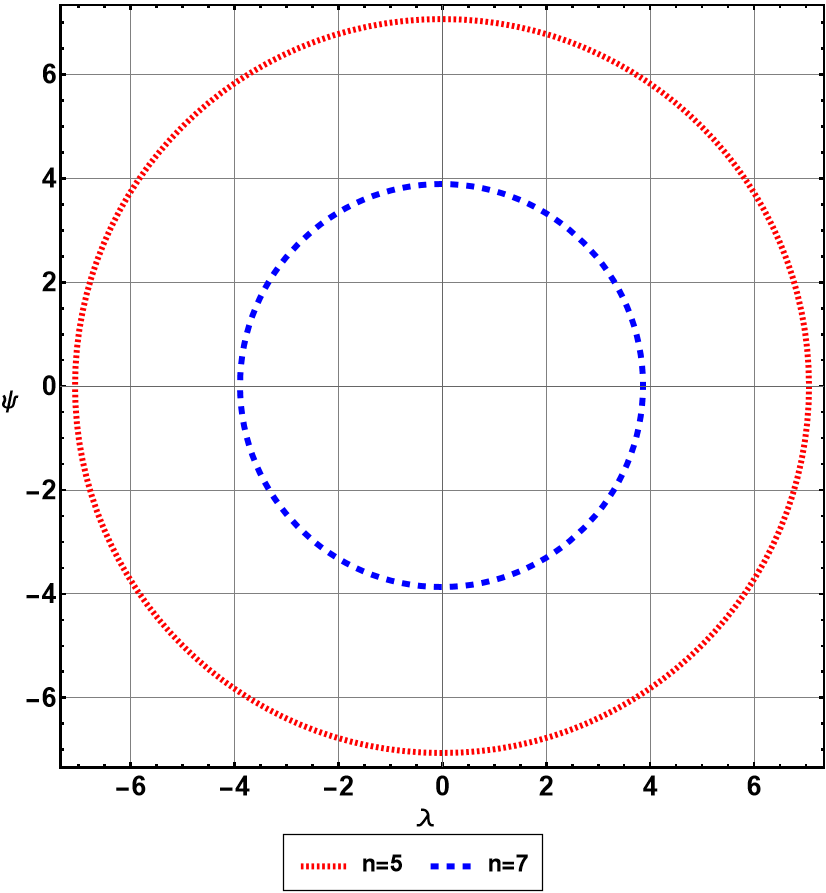}}
\,\,\,
\subfloat[\label{Fig10e} for $\Lambda=-0.10$]{\includegraphics[width=0.29\textwidth]{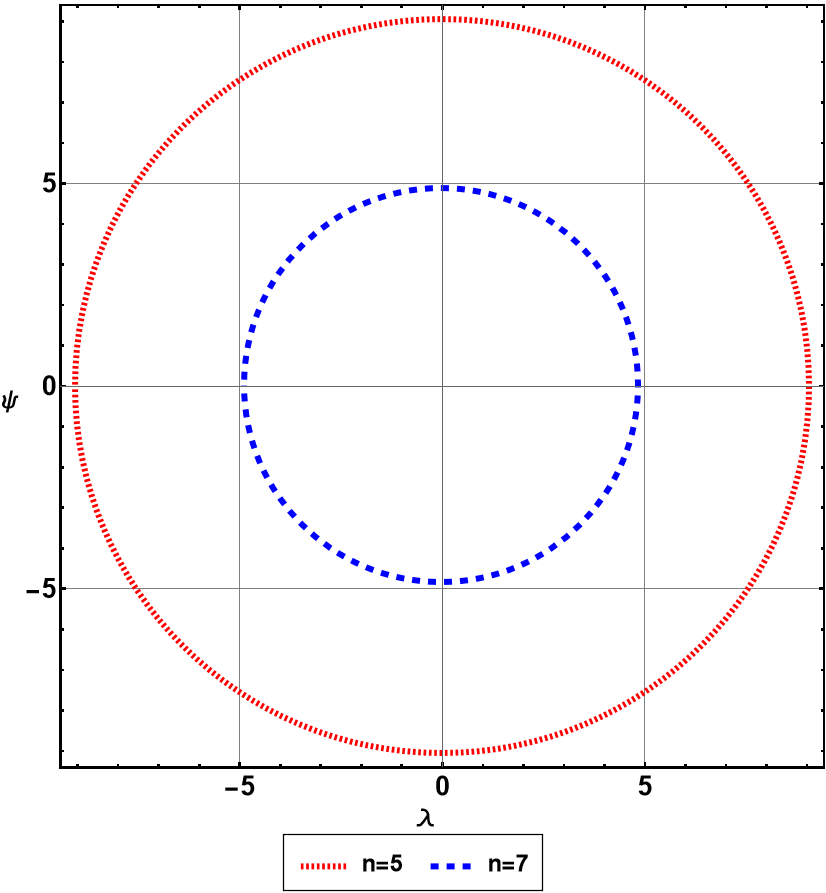}}
\caption{\label{Fig10}\small{\emph{Geometrical shape of the shadow of the higher-dimensional ALAdS charged black hole in celestial plane with $M=1$ using the set $(\alpha=0.01,\,\gamma=0.51)$.}}}
\end{figure}

Using the data collected in Table \ref{Table2}, we illustrate of the shadow shapes in celestial coordinates of the higher-dimensional ALAdS black hole in EHM gravity in Fig. \ref{Fig10}. Each plot in Fig. \ref{Fig10} is illustrated for a fixed value of $\Lambda$. We see from Fig. \ref{Fig10} that when the cosmological constant $\Lambda$ is fixed, the radius of shadow circles of the ALAdS charged black hole (like ALF one) decreases with increasing the number of dimensions.

\begin{figure}[htb]
\centering
\subfloat[\label{Fig11a} for $n=4$]{\includegraphics[width=0.284\textwidth]{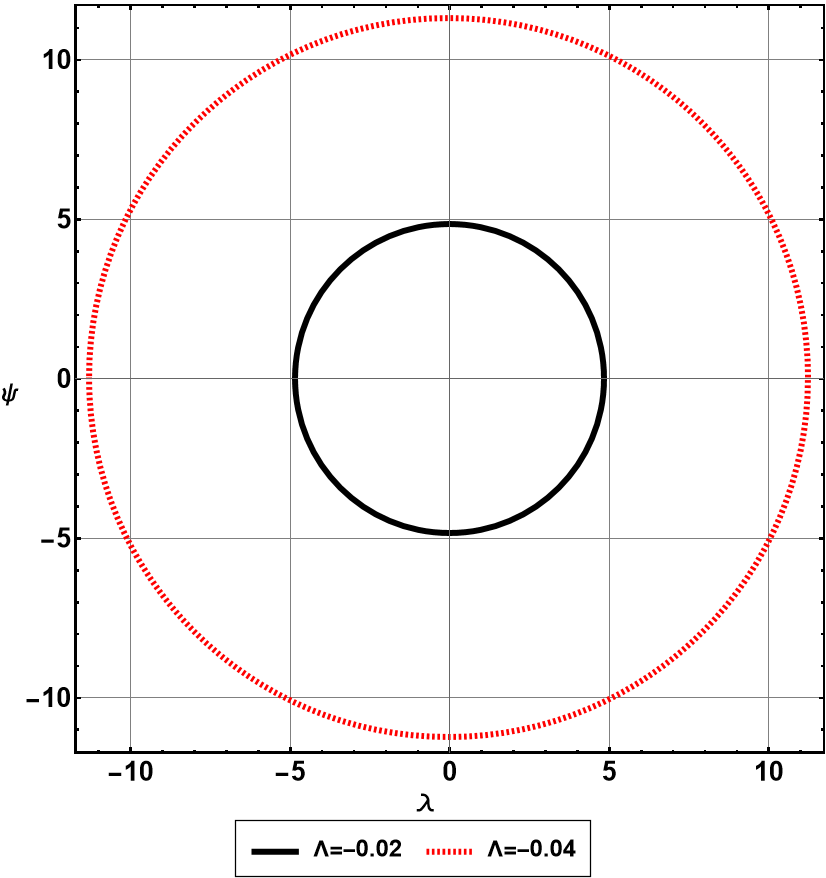}}
\,\,\,
\subfloat[\label{Fig11b} for $n=5$]{\includegraphics[width=0.271\textwidth]{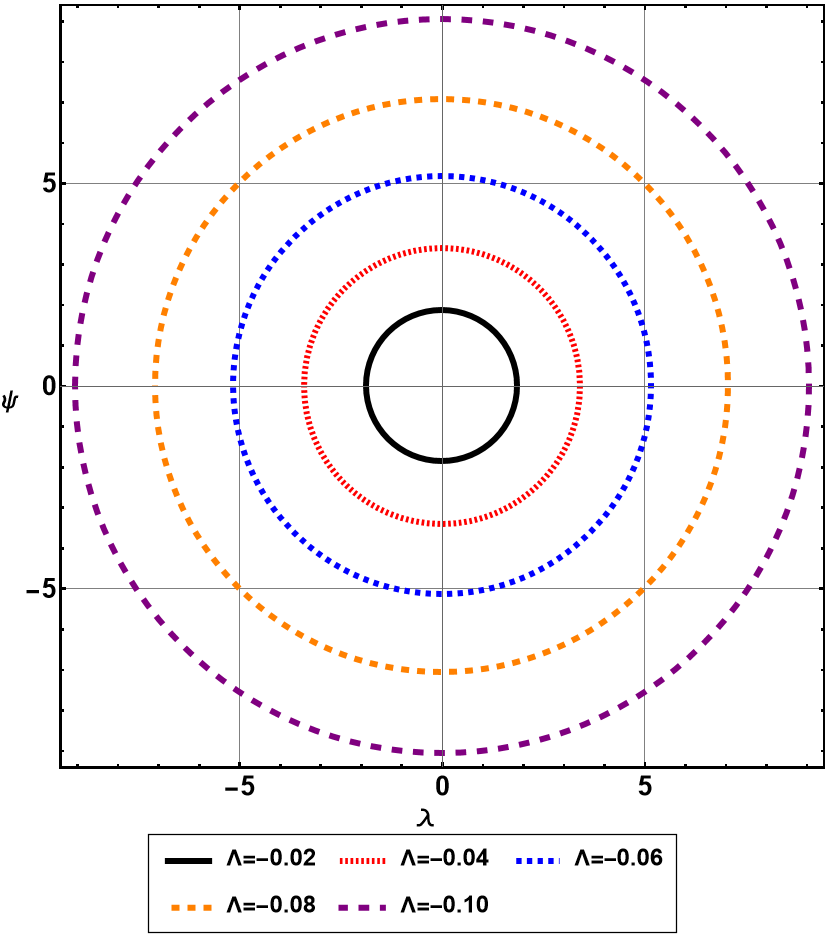}}
\,\,\,
\subfloat[\label{Fig11c} for $n=7$]{\includegraphics[width=0.271\textwidth]{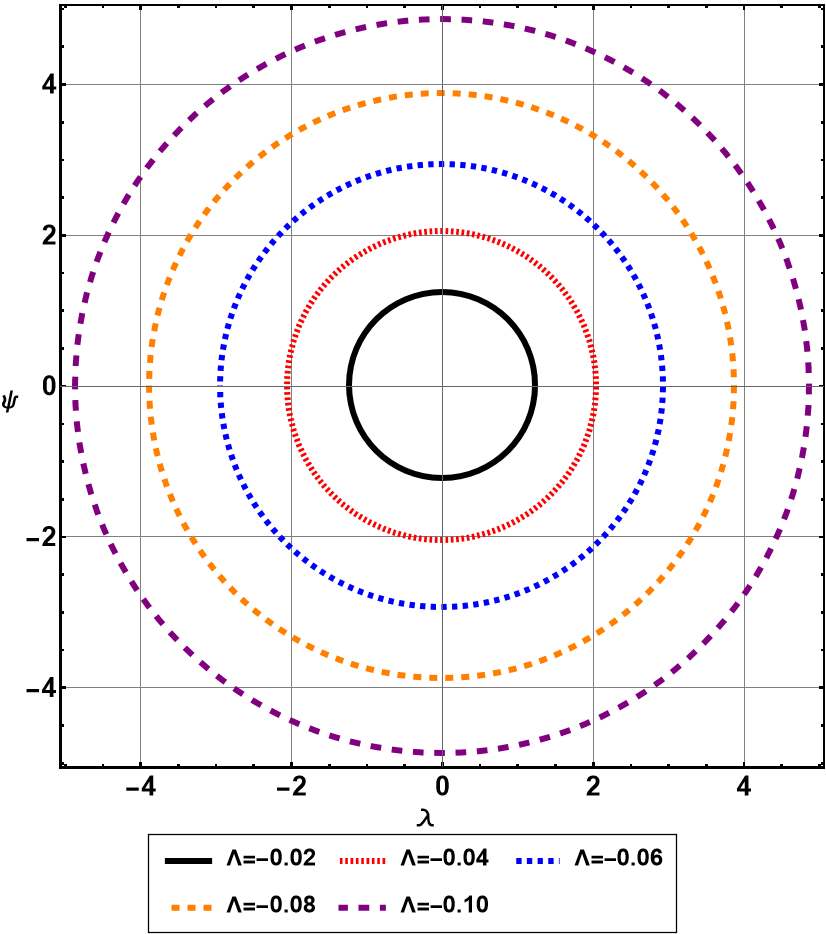}}
\caption{\label{Fig11}\small{\emph{Geometrical shape of the shadow of the higher-dimensional ALAdS charged black hole in celestial plane with $M=1$ using the set $(\alpha=0.01,\,\gamma=0.51)$.}}}
\end{figure}

Now by fixing the number of dimensions $n$, we plot the shadow circles of the ALAdS charged black hole with extra dimensions in EHM gravity in Fig. \ref{Fig11} for different values of $\Lambda$. From Fig. \ref{Fig11}, we can obviously find that decreasing the value of the negative cosmological constant (i.e., increasing its absolute value) results in remarkably growing the radius of the shadow circles of the ALAdS charged black hole. This means that turning off the cosmological constant yields smaller shadow sizes. Consequently, for the higher-dimensional charged ALAdS black hole in EHM gravity, from Figs. \ref{Fig10} and \ref{Fig11}, we can see that the impact of the extra dimensions (cosmological constat) on the shadow of the black hole is to reduce (amplify) its size.

\subsubsection{Energy emission rate}

One can insert $r_{eh}$ values from Table \ref{Table2} into Eq. \eqref{Htemalads} to find numerical values of the Hawking temperature of the ALAdS extra dimensional black hole in EHM gravity. Then, the numerical values of $\sigma_{lim}$ in Eq. \eqref{sigmalim} for different $\Lambda$ and $n$ from the values of the shadow radius $\sqrt{\eta+\xi^{2}}$ of the black hole should be found. Consequently, one can find the expressions of the energy emission rate in terms of different values of $\Lambda$ and $n$ associated with the higher-dimensional ALAdS black hole in EHM gravity by inserting the Hawking temperature and $\sigma_{lim}$ values into Eq. \eqref{radirate}.

\begin{figure}[H]
\centering
\subfloat[\label{Fig12a} for $\Lambda=-0.02$]{\includegraphics[width=0.49\textwidth]{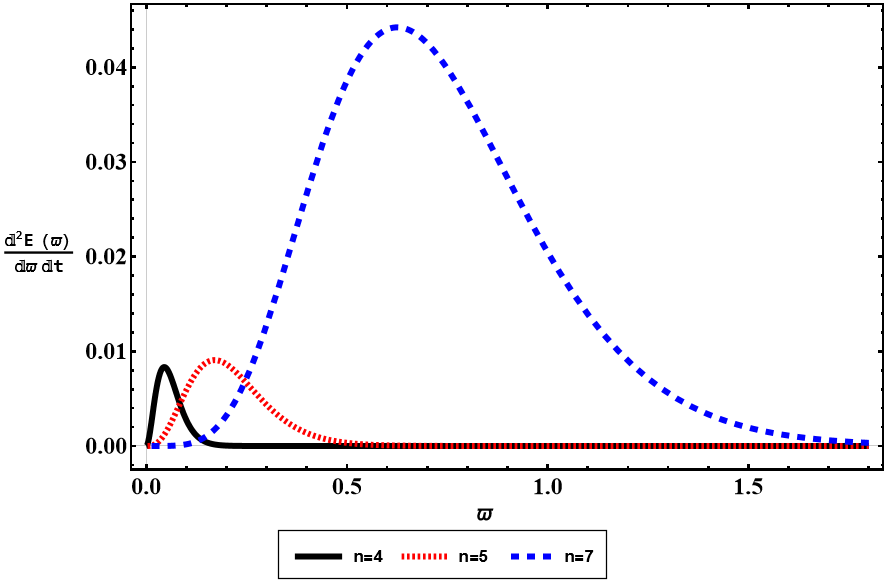}} \\
\subfloat[\label{Fig12b} for $n=5$]{\includegraphics[width=0.49\textwidth]{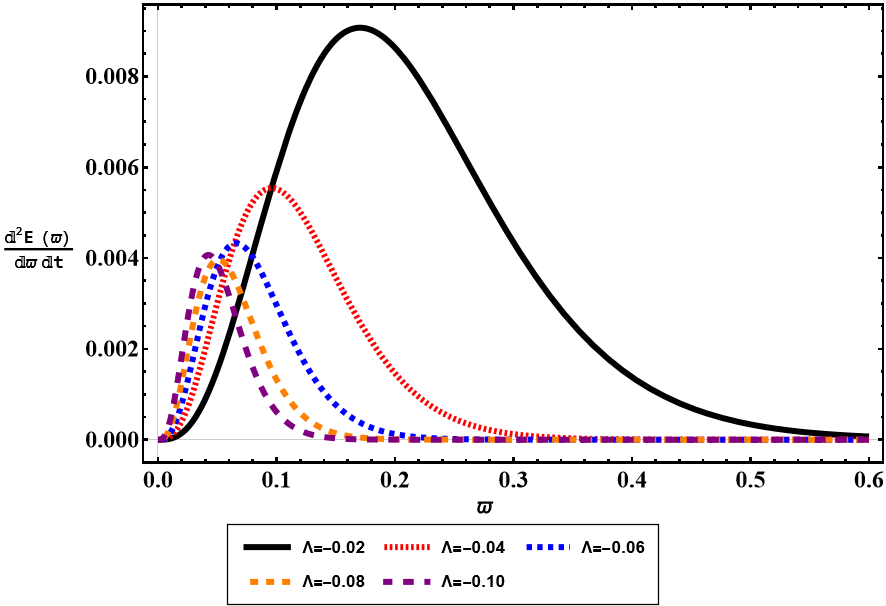}}
\,\,\,
\subfloat[\label{Fig12c} for $n=7$]{\includegraphics[width=0.49\textwidth]{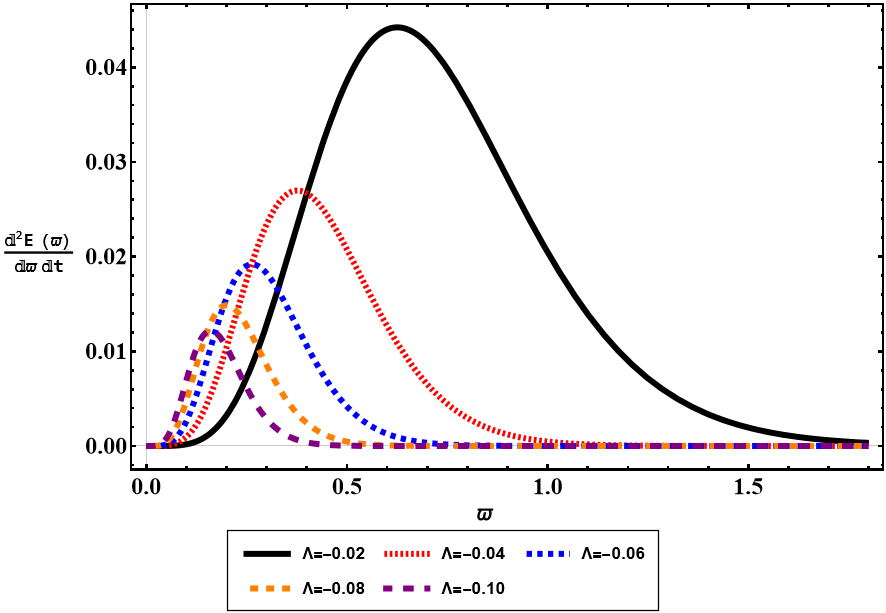}}
\caption{\label{Fig12}\small{\emph{The energy emission rate as a function of $\varpi$ for the higher-dimensional ALAdS charged black hole in EHM gravity for different values of $n$ and $\Lambda$.}}}
\end{figure}

Fig. \ref{Fig12} is the illustration of the energy emission rate for the charged ALAdS higher-dimensional black hole in EHM gravity in terms of the emission frequency $\varpi$. In Fig. \ref{Fig12a} the behavior of the energy emission rate is shown for the fixed value of the cosmological constant $\Lambda=-0.02$ associated with $n=4, 5$, and $7$. Additionally, Figs. \ref{Fig12b} and \ref{Fig12c} are related to $n=5$ and $n=7$, respectively. From Fig. \ref{Fig12a} we see that for a fixed value of $\Lambda$, growing $n$ results in significantly increasing the energy emission rate of the ALAdS charged black hole with extra dimensions. Thus, like ALF charged black hole, we found out that extra dimensions accelerate the evaporation of the ALAdS charged black hole with higher dimensions. Moreover, From Figs. \ref{Fig12b} and \ref{Fig12c} we see that for a fixed value of number of dimensions $n$, decreasing $\Lambda$ (i.e., increasing its absolute value) results in reducing the energy emission rate of the ALAdS charged black hole with extra dimensions. This means that turning off the cosmological constant leads to amplify the energy emission rate. Also, the curves of the energy emission of the black hole in Fig. \ref{Fig12c} associated with $n=7$ have larger values than the corresponding ones in Fig. \ref{Fig12b} for $n=5$. Consequently, the impact of the cosmological constant is to reduce the energy emission rate of the charged ALAdS higher-dimensional black hole, which results in decelerate its evaporation while the effect of extra dimensions is to accelerate it.

\subsubsection{Deflection angle}

We apply Eqs. \eqref{fralads}-\eqref{hrq} and \eqref{hbarrodd} for odd $n$'s for simplicity on Eqs. \eqref{optmet2} and \eqref{gauoptcur1} to gain the Gaussian optical curvature for the higher-dimensional ALAdS charged black hole. The Gaussian optical curvature of the black hole up to the first order in source mass and cosmological constant and second order in electric charge and again second order in coupling constants $\gamma$ and $\alpha$, can be found as follows
\begin{equation}\label{gauoptcur3}
\begin{split}
K\approx\frac{-8(n-3)\pi^{1-n}r^{-2n}}{\gamma^{2}(n-3)^{4}(n-2)^{5}(n-1)(n+1)} & \bigg\{\alpha^{2}M(n-3)(n-2)^{2}(n^{2}-1)(n-4)(5n-7)\pi^{\frac{n+1}{2}}r^{n+5}\Gamma\left[\frac{n-1}{2}\right]\\
& +6\alpha\gamma\Lambda\pi^{n-1}(n-3)(n-2)^{3}(n+1)(2n-3)r^{2-2n}\\
& +540\pi^{2}Q^{2}r^{4}\gamma^{2}\left(\Gamma\left[\frac{n-1}{2}\right]\right)^{2}\bigg\}\,.
\end{split}
\end{equation}
Furthermore, the surface element of the optical metric \eqref{optmet2} for the higher-dimensional ALAdS charged black hole in EHM gravity corresponding with the metric coefficients \eqref{fralads}-\eqref{hrq} and \eqref{hbarrodd} for odd $n$'s can be found approximately as the same as Eq. \eqref{surele1}. Now, by inserting Eqs. \eqref{surele1} and \eqref{gauoptcur3} into the deflection angle expression \eqref{defang}, one can get approximately the deflection angle of the higher-dimensional ALAdS black hole in EHM gravity as follows
\begin{equation}\label{defang2}
\begin{split}
\Theta & =-\int_{0}^{\pi}\int_{\frac{\xi}{\sin[\phi]}}^{\infty}KdS\\
& = -\int_{0}^{\pi}\int_{\frac{\xi}{\sin[\phi]}}^{\infty}Krdrd\phi\\
& \approx\frac{\xi^{4-4n}}{(n-3)^{2}(n-2)^{4}}\Bigg\{\frac{56\alpha^{2}M(n-5)(n-3)(n-2)\pi^{\frac{4-n}{2}}\xi^{3n+3}\Gamma\left[\frac{n-6}{2}\right]}{\gamma^{2}(n-7)}
+\frac{3\sqrt{\pi}\alpha\Lambda(n-2)\Gamma\left[\frac{4n-3}{2}\right]}{\gamma(n-1)^{3}\Gamma[2n-4]}\\
& +\frac{540\pi^{\frac{7-2n}{2}}Q^{2}\xi^{2n+2}\left(\Gamma\left[\frac{n-3}{2}\right]\right)^{2}\Gamma\left[\frac{2n-5}{2}\right]}{(n+1)\Gamma[n]}\Bigg\}\,.
\end{split}
\end{equation}

\begin{figure}[htb]
\centering
\subfloat{\includegraphics[width=0.5\textwidth]{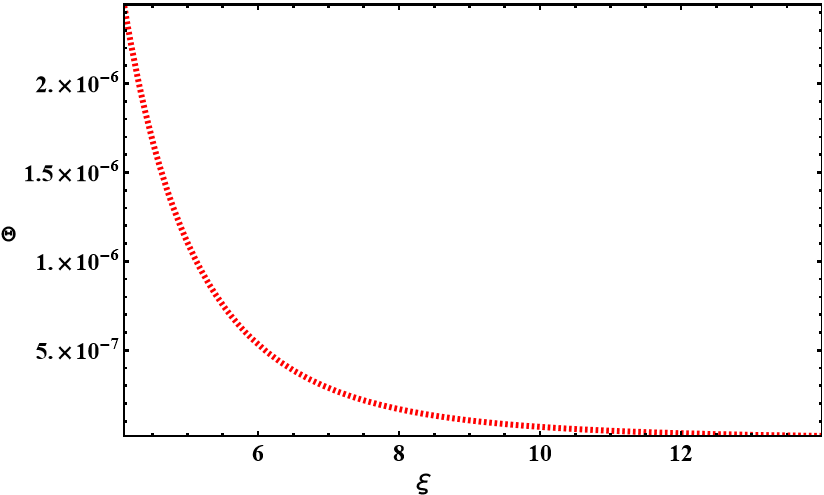}}
\caption{\label{Fig13}\small{\emph{The behavior of deflection angle of the higher-dimensional ALAdS black hole in EHM gravity in terms of $\xi$ for $n=5$ and $\Lambda=-0.02$.}}}
\end{figure}
Fig. \ref{Fig13} is the illustration of the deflection angle of the higher-dimensional ALAdS black hole in EHM gravity versus the impact parameter $\xi$ with respect to $\Lambda=-0.02$ and $n=5$ for simplicity. From Fig. \ref{Fig13}, we see that reducing the impact parameter $\xi$ again results in increasing the deflection angle of the black hole.

\subsubsection{Constraints from EHT observations of M87*}

Now we want to compare the shadow radius of the higher-dimensional ALAdS charged black hole in EHM gravity with the shadow size of M87* supermassive black hole captured by EHT in Eq. \eqref{M87} to constrain the cosmological constant values and coupling constants of the EHM theory.

\begin{figure}[htb]
\centering
\subfloat[\label{Fig14a} for the set $(\alpha=0.01,\,\gamma=0.51)$]{\includegraphics[width=0.49\textwidth]{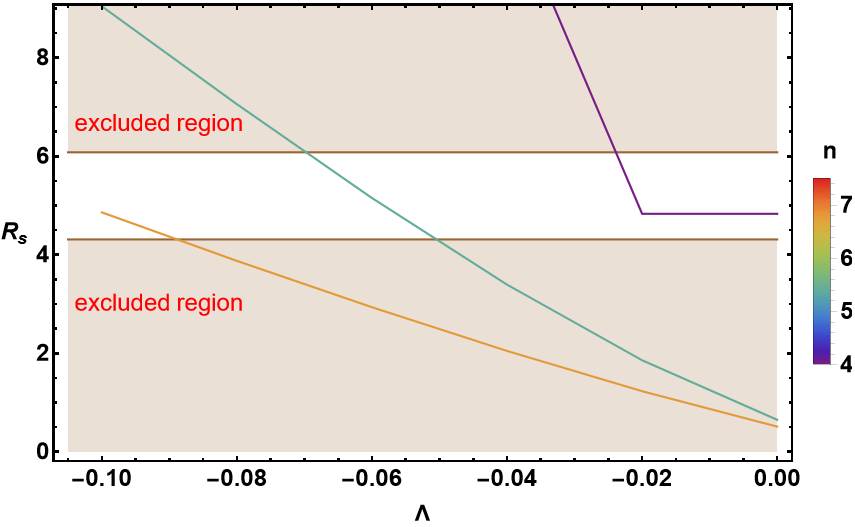}}
\,\,\,
\subfloat[\label{Fig14b} for the set $(\alpha=0.015,\,\gamma=0.81)$]{\includegraphics[width=0.49\textwidth]{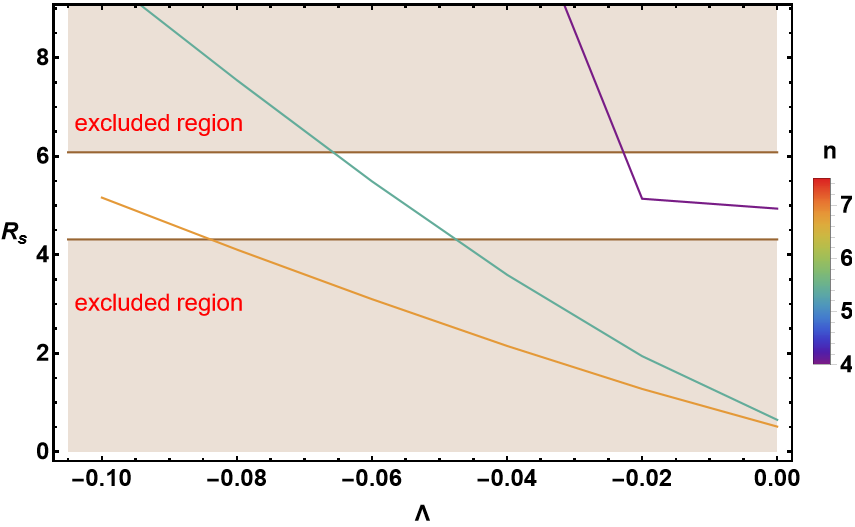}} \\
\subfloat[\label{Fig14c} for the set $(\alpha=0.015,\,\Lambda=-0.10)$]{\includegraphics[width=0.49\textwidth]{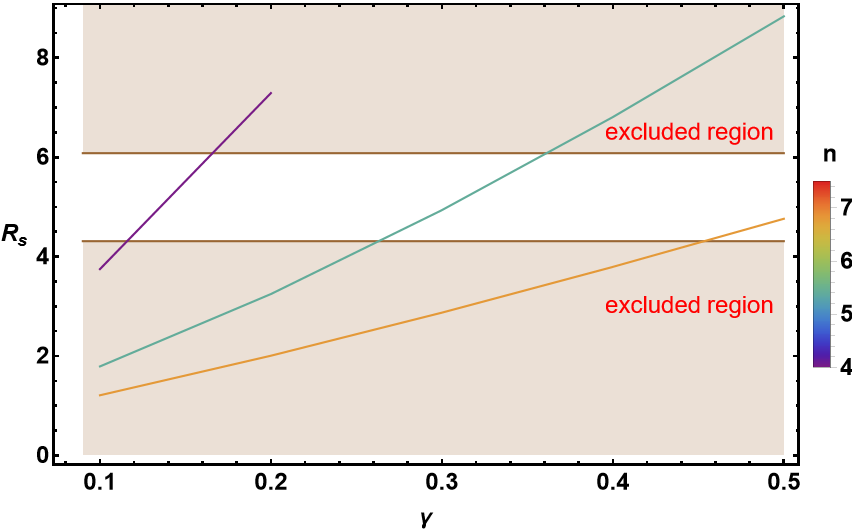}}
\,\,\,
\subfloat[\label{Fig14d} for the set $(\gamma=0.51,\,\Lambda=-0.10)$]{\includegraphics[width=0.49\textwidth]{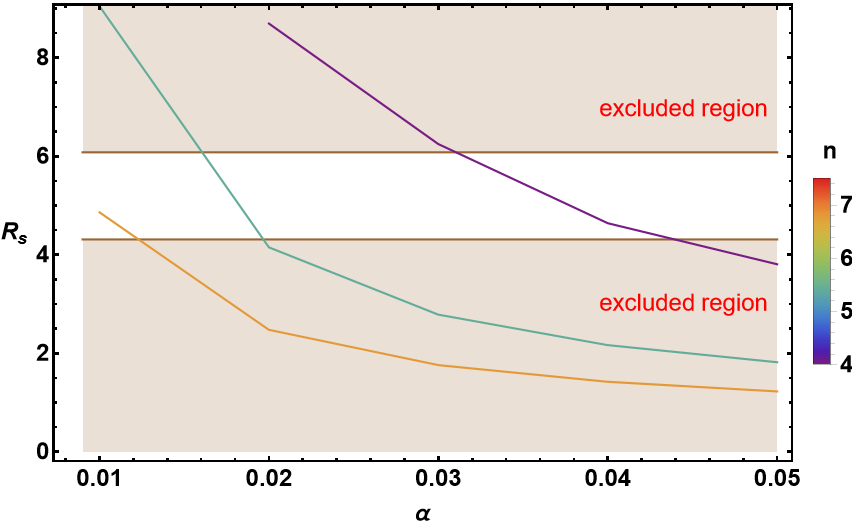}}
\caption{\label{Fig14}\small{\emph{The shadow radius of the higher-dimensional ALAdS charged black hole in EHM gravity in comparison with the shadow size of M87* captured via EHT within $1$-$\sigma$ confidence level. The brown (shaded) areas are the excluded regions, which are inconsistent with the observations of EHT while the white (unshaded) region is the $1$-$\sigma$ confidence level of EHT data.}}}
\end{figure}
Figure \ref{Fig14} shows the behavior of the shadow radius of the higher-dimensional ALAdS charged black hole in EHM gravity in comparison with the shadow radius of M87* given by EHT within $1$-$\sigma$ uncertainties as seen in Eq. \eqref{M87}. In Fig. \ref{Fig14}, the white (unshaded) region indicates the $1$-$\sigma$ confidence level while the brown (shaded) areas are the excluded regions, which are inconsistent with the observations of EHT related to the shadow radii of M87*. In Figs. \ref{Fig14a} and \ref{Fig14b} utilizing the sets $(\alpha=0.01,\,\gamma=0.51)$ and $(\alpha=0.015,\,\gamma=0.81)$, respectively, the comparison between the shadow radius of the higher-dimensional ALAdS charged black hole and M87* is shown versus the cosmological constant. From Fig. \ref{Fig14a} we see that for $n=4$, the shadow radius of the black hole is in consistence with M87* shadow for $0<\Lambda<-0.025$ while for $n=5$ and $n=7$ such a consistency can be seen in $-0.05<\Lambda<-0.07$ and $-0.09<\Lambda$, respectively. Comparing Fig. \ref{Fig14a} and Fig. \ref{Fig14b} shows that these ranges shift a bit towards larger values of $\Lambda$ while the values of the shadow radius of the black hole experiences a tiny amplification by increasing $\alpha$ and $\gamma$ in Fig. \ref{Fig14b}. It should be noted that Figs \ref{Fig14a} and \ref{Fig14b} show that omitting the cosmological constant leads to reduction of the shadow radius of the black hole. Moreover, Fig. \ref{Fig14c} is for comparison between the shadow radii of the higher-dimensional ALAdS charged black hole and M87* versus the coupling constant $\gamma$ with respect to the set $(\alpha=0.015,\,\Lambda=-0.10)$. From Fig. \ref{Fig14c}, we see that for $n=4$, the shadow radius of the black hole is compatible with M87* shadow in the range $0.12<\gamma<0.17$ while for $n=5$ and $n=7$ such a compatibility appears in $0.26<\gamma<0.36$ and $0.46<\gamma$, respectively. Also, Fig. \ref{Fig14c} indicates that increasing the coupling constant $\gamma$ leads to amplification of the shadow radius of the black hole. Additionally, Fig. \ref{Fig14d} is for comparing the shadow radius of the higher-dimensional ALAdS charged black hole and M87* versus coupling constant $\alpha$ using the set $(\gamma=0.51,\,\Lambda=-0.10)$. In Fig. \ref{Fig14d}, we see that for $n=4$, the shadow radius of the black hole is compatible with M87* shadow in the interval $0.03<\alpha<0.044$ while for $n=5$ and $n=7$ such a compatibility appears in $0.016<\alpha<0.02$ and $\alpha<0.012$, respectively. We see from Fig. \ref{Fig14d} that increasing the coupling constant $\alpha$ results in reducing the shadow size of the black hole. The key point here is that from Fig. \ref{Fig14} one can expect that the extra dimensions, especially $n=5$ can be apparently observed from the shadow of black holes captured by EHT thanks to the presence of the cosmological constant in the EHM theory.

\section{Summary and Conclusions}\label{consum}

In this study, according to string theory, braneworld models, and AdS/CFT correspondence, we motivated to take into account the higher-dimensional ALF and ALAdS charged black hole solutions of the EHM theory to investigate the behaviors of the corresponding shadow and deflection angle. Our main goal was to discover how extra dimensions and the other parameters of the theory affect the shadow of the black holes. To do this, we first provided the required general formalism to study the shadow behavior of these higher-dimensional black holes utilizing the Hamilton-Jacobi approach and Carter method to formulate the null geodesics around them and derive the corresponding effective potentials. Next, we introduced the celestial coordinates to specify the shadow shape of the higher-dimensional black holes on the observer's sky. We also estimated the energy emission rate and deflection angle formulas in the higher-dimensional scenario. Additionally, we introduced the black hole shadow observables including shadow size and distortion, as well as shadow area and oblateness proposed by Hioki-Maeda and Kumar-Ghosh proposals, respectively. Then, employing the constructed framework, we studied the shadow behavior, deflection angle, and energy emission rate of the ALF and ALAdS charged black holes in EHM gravity with extra dimensions. We computed and analyzed the significant impacts of the electric charge, cosmological constant, and extra dimensions on the shadow, deflection angle, and energy emission rate of the black holes within the setup. Moreover, we constrain these parameters by comparing the shadow size of M87* from EHT observations with the shadow radius of the higher-dimensional ALF and ALAdS charged black holes.

For the higher-dimensional charged ALF case, we discovered that for a fixed value of the electric charge $Q$, the shadow size of the black hole decreases with increasing the number of extra dimensions $n$. Also, when the electric charge value increases with a fixed $n$, the shadow size of the black hole again decreases, whereas the effect of the electric charge in comparison with extra dimensions on the shadow of charged higher-dimensional ALF black hole is suppressible. Also, we saw that for a fixed value of $Q$, the energy emission rate of the ALF charged black hole with extra dimensions extremely increases by growing $n$. Also, growing the electric charge increases the energy emission rate, but its effect can be eliminated. Therefore, we found that extra dimensions accelerate the evaporation of the ALF black hole in EHM gravity with higher dimensions. Then, using the Gauss-Bonnet theorem, we have calculated the leading terms of the deflection angle in the weak-limit approximation. We have discussed the impact of charge and the extra dimensions on this optical quantity. It was obvious that for a fixed value of the electric charge, the deflection angle of the ALF black hole in EHM gravity with extra dimensions reduces by growing the number of dimensions. Also, for a fixed value of $n$, the deflection angle of the black hole decreases by increasing the electric charge $Q$. However, the effect of the electric charge is dominated by the impact of extra dimensions on the deflection angle of the black hole. Furthermore, by comparing the shadow radius of the black hole with M87* shadow released by EHT, we observed that only the shadow of four dimensional ALF charged black hole with $0\leq Q<1.8$ lies in the $1$-$\sigma$ uncertainties of EHT data.

On the other hand, for the higher-dimensional ALAdS charged black hole in the EHM gravity we observed that when the negative cosmological constant $\Lambda$ is fixed, the radii of shadow circles of the higher-dimensional ALAdS charged black hole decrease by increasing the number of extra dimensions $n$. However, for a fixed $n$, the shadow radius of the higher-dimensional ALAdS charged black hole increases by decreasing the negative cosmological constant (i.e., increasing its absolute value). Also, we found that for a fixed value of $\Lambda$, the energy emission rate of the ALAdS charged black hole with extra dimensions extremely increases by growing $n$, whereas for a fixed $n$, the energy emission rate of the black hole decreases by decreasing the negative cosmological constant (i.e., increasing its absolute value). Hence, we found that extra dimensions and negative cosmological constant accelerate the evaporation of the ALAdS charged black hole with higher dimensions. Moreover, we observed that increasing the coupling constant $\gamma$ of the EHM gravity leads to amplify the shadow radius of the black hole whereas increasing the coupling constant $\alpha$ of the theory results in reducing the shadow size of the black hole. Surprisingly, by comparing the shadow of M87* captured by EHT with the shadow of the black hole, we proved that the four, five, and seven dimensional ALAdS charged black hole are compatible with EHT data thanks to the presence of the negative cosmological constant.

In summary, we can came to conclusion that the shadows of higher-dimensional ALF and ALAdS charged black holes in EHM theory are characterized by the extra dimensions in addition to their parameters. In this regard, the extra dimensions within EHM theory affect the shadows of the black holes by reducing their size, significantly. On the other hand, owing to the existence of the negative cosmological constant within EHM theory, we concluded that it seems possible to detect the effects of the extra dimensions via EHT. The key point here is that from Fig. \ref{Fig14}, one can expect that the extra dimensions, especially $n=5$ can be apparently observed from the shadow of black holes captured by EHT thanks to the presence of the cosmological constant in the EHM theory. These outcomes may lead to the possibility of testing the higher-dimensional charged black hole solutions of EHM gravity by employing astrophysical observations.

\begin{acknowledgments}

The authors would like to thank Milad Hajebrahimi for fruitful comments and discussions. Also, the authors appreciate the respectful referees for carefully reading the manuscript and their insightful comments which boosted the quality of the paper, considerably.

\end{acknowledgments}

\end{document}